\documentclass[11pt,a4paper]{article}
\pdfoutput=1

\newcommand{\cor}[1] {\textcolor{black}{#1}}

\usepackage{
    jcappub,
    bbm,
    mathtools,
    microtype
}

\begin{document}

\title{Resummed Kinetic Field Theory: Using Mesoscopic Particle Hydrodynamics to Describe Baryonic Matter in a Cosmological Framework}

\author[a,b,c]{Daniel Geiss,}
\author[a,d]{Robert Lilow,}
\author[a]{Felix Fabis,}
\author[a]{Matthias Bartelmann}

\affiliation[a]{Heidelberg University, Zentrum f\"ur Astronomie, Institut f\"ur Theoretische Astrophysik, Philosophenweg 12, 69120 Heidelberg, Germany}
\affiliation[b]{Institut f\"ur Theoretische Physik, Leipzig University, 04103 Leipzig, Germany}
\affiliation[c]{Max Planck Institute for Mathematics in the Sciences, 04103 Leipzig, Germany}
\affiliation[d]{Department of Physics, Technion, Haifa 3200003, Israel}

\emailAdd{daniel.geiss@mis.mpg.de}
\emailAdd{rlilow@campus.technion.ac.il}
\emailAdd{felix.fabis@posteo.de}
\emailAdd{bartelmann@uni-heidelberg.de}

\abstract{\cor{We present a new analytical description of baryonic matter in a cosmological framework, using the formalism of \textquoteleft Kinetic Field Theory\textquoteright\ (KFT) -- a statistical field theory approach to structure formation based on the dynamics of classical particles. So far, only the purely gravitational dynamics of dark matter had been considered in KFT, but an accurate description of cosmic structure formation requires to also take into account the baryonic gas dynamics. In this paper, we propose to achieve this by incorporating an effective mesoscopic particle model of hydrodynamics into the recently developed framework of Resummed KFT. Our main result is the baryonic density contrast power spectrum computed to lowest perturbative order, assuming a simplified isothermal fluid model. Compared to the spectrum of dark matter, the baryonic spectrum shows a suppression of power as well as an oscillatory behaviour associated with sound waves on scales smaller than the Jeans length. We further compare our result to the linear spectrum of an isothermal fluid obtained from Eulerian perturbation theory (EPT), finding good quantitative agreement within the approximations we made in the EPT calculation. A subsequent paper will resolve the problem of coupling both dark and baryonic matter, to gain a full model of cosmic matter. Applying the mesoscopic particle approach to more general ideal or viscous fluids will also be subject of upcoming work.}}

\arxivnumber{1811.07741}


\maketitle

\flushbottom

%%%%%%%%%%%%%%%%%%%%%%%%%%%%%%%%%%%%%%%%
\section{Introduction}
Most of our current knowledge about cosmic structure formation is based on numerical $N$-body simulations, which allow to follow the highly nonlinear evolution of those structures but offer only little insight into the underlying physics. \cor{For analytical and semi-analytical approaches, on the other hand, which allow to gain a deeper understanding of these physical processes, describing structure formation beyond the linear and mildly nonlinear regime has proven to be extremely challenging (see \cite{bouchet_perturbative_1995,valageas_new_2004,crocce_renormalized_2006,crocce_memory_2006,matarrese_resumming_2007,matsubara_resumming_2008,pietroni_coarse-grained_2012,anselmi_nonlinear_2012,carrasco_effective_2012,blas_cosmological_2014,porto_lagrangian-space_2014,senatore_ir-resummed_2015} for a selection of different approaches and \cite{bernardeau_large-scale_2002} for an extensive review).}
\par
\cor{Aiming at overcoming these difficulties,} Bartelmann et al. \cite{bartelmann_microscopic_2016,bartelmann_kinetic_2017,bartelmann_analytic_2017} lately developed a new analytical approach based on the pioneering ideas of Das and Mazenko \cite{mazenko_fundamental_2010,mazenko_smoluchowski_2011,das_field_2012,das_newtonian_2013}. The so-called `Kinetic Field Theory' (KFT) mimics the concepts of $N$-body simulations by considering the dynamics of single particles \cor{in phase-space. All collective quantities such as correlators of the density or velocity fields} can compactly be obtained from a generating functional by applying appropriate operators.
\par
So far, this formalism has been used to describe a universe consisting of only dark matter, whose interactions are purely \cor{gravitational} and thus only provide an accurate description of matter on large scales. To enable an accurate treatment of small-scale phenomena, a consideration of baryonic degrees of freedom becomes necessary. \cor{While there has been some work on analytical descriptions of baryonic matter, e.g.~\cite{nusser_analytic_2000,matarrese_growth_2002,shoji_third-order_2009}, they are so far mostly investigated using numerical simulations, see e.g.~\cite{rudd_effects_2008,angulo_how_2013,vogelsberger_introducing_2014,valkenburg_accurate_2017,fidler_new_2019}.} The purpose of this work is to find a practicable way to extend the existing KFT formalism to a model describing baryonic matter in a cosmological framework, while a discussion of a coupled theory including dark matter will be given in a subsequent paper. 
\par
A first guess for a way to tackle this issue would be to consider the existing model of KFT -- which describes dark matter -- and replace the interaction potential by another, appropriate one describing the interactions between baryons. But there is no obvious way how to choose such a potential. Making the assumption that baryonic matter on large scales is well described by the hydrodynamical equations, Viermann et al. \cite{viermann_model_2018} instead proposed a model based on (effective) mesoscopic particles, which is inspired by the ideas of Smoothed Particle \cor{Hydrodynamics \cite{monaghan_smoothed_1992}}. We adopt the underlying picture of a mesoscopic particle and present an alternative derivation of the equations of motion for an ideal gas which verifies the findings of \cite{viermann_model_2018}. \cor{It allows some better physical interpretation and enables us to to generalize the results to an expanding space-time.} The inclusion of the equations of motion of a system of mesoscopic particles in the KFT language is what we call `Mesoscopic Particle Hydrodynamics' (MPH).
\par
In \cite{viermann_model_2018} MPH was investigated using a perturbative expansion in orders of the interactions between mesoscopic particles. While low-order calculations already exhibited characteristic fluid-like behaviour, they also indicated that any perturbative expansion to finite order in the mesoscopic interactions is most likely insufficient. To overcome these limitations, we show how MPH can be incorporated into the Resummed KFT (RKFT) framework introduced by Lilow et al. \cite{lilow_resummed_2019}. \cor{The leading-order result in RKFT already captures significant hydrodynamical effects that in the standard perturbative expansion of KFT could only be described at infinitely high order in the mesoscopic interactions.} We confirm the validity of this resummation by comparing our results to \cor{those obtained from Eulerian perturbation theory (EPT)}.
\par
Our focus lies on the investigation of cosmic structure formation, hence the main result of this work is the calculation of a purely baryonic density power spectrum in a cosmological framework. For this, we approximate the baryons as an isothermal fluid, since this simplifies the treatment in KFT while still capturing the main qualitative differences to the collisionless dynamics of dark matter. The extension to more realistic baryonic fluid dynamics will be discussed in upcoming work. In this paper, we further restrict ourselves to an Einstein-de Sitter (EdS) model, which corresponds to a flat matter-only universe. This model poses a good description \cor{during the early times after CMB decoupling}.
\par
This work is structured as follows: In \autoref{Section MPH}, the idea of a mesoscopic particle is \cor{revisited} and its equations of motion are derived based on a projection procedure applied to the hydrodynamical equations. Considering an isothermal fluid, we show how the equations of motion can be generalized to an expanding universe. \cor{After reviewing the general KFT formalism and the resummation procedure according to \cite{lilow_resummed_2019}, we apply it to the mesoscopic interactions of isothermal MPH in \autoref{Section KFT}. We calculate the density contrast power spectrum at lowest perturbative level within RKFT for an EdS universe containing only baryons and compare this to the spectrum of dark matter.} In \autoref{Section EPT}, we show how baryonic matter may be treated in Eulerian perturbation theory \cite{bernardeau_large-scale_2002} and compare the density power spectrum resulting from the linearized hydrodynamical equations with our findings from MPH. In \autoref{Section Conclusion}, the most important aspects are summarized and an outlook is given.

%%%%%%%%%%%%%%%%%%%%%%%%%%%%%%%%%%%%%%%%
\section{Mesoscopic Particle Hydrodynamics}
\label{Section MPH}

%%%%%%%%%%%%%%%%%%%%
\subsection{The Idea of a Mesoscopic Particle}
\label{Section Idea of MPH}
\cor{One of the techniques commonly used to describe fluid dynamics in numerical simulations is Smoothed Particle Hydrodynamics \cite{monaghan_smoothed_1992}. Its} main idea is that every function of space can be approximated by a finite sum of spatially localized contributions, interpreted as contributions from individual particles. With this ansatz the hydrodynamical equations can be reformulated as a model of effective particles. Viermann et al. \cite{viermann_model_2018} adopted this ansatz to make it applicable in the context of KFT. In the following we \cor{revisit} the basic idea of a mesoscopic particle and present an alternative derivation of the equations of motion.
\par
\begin{figure}
	\centering
	\includegraphics[width=0.6\textwidth]{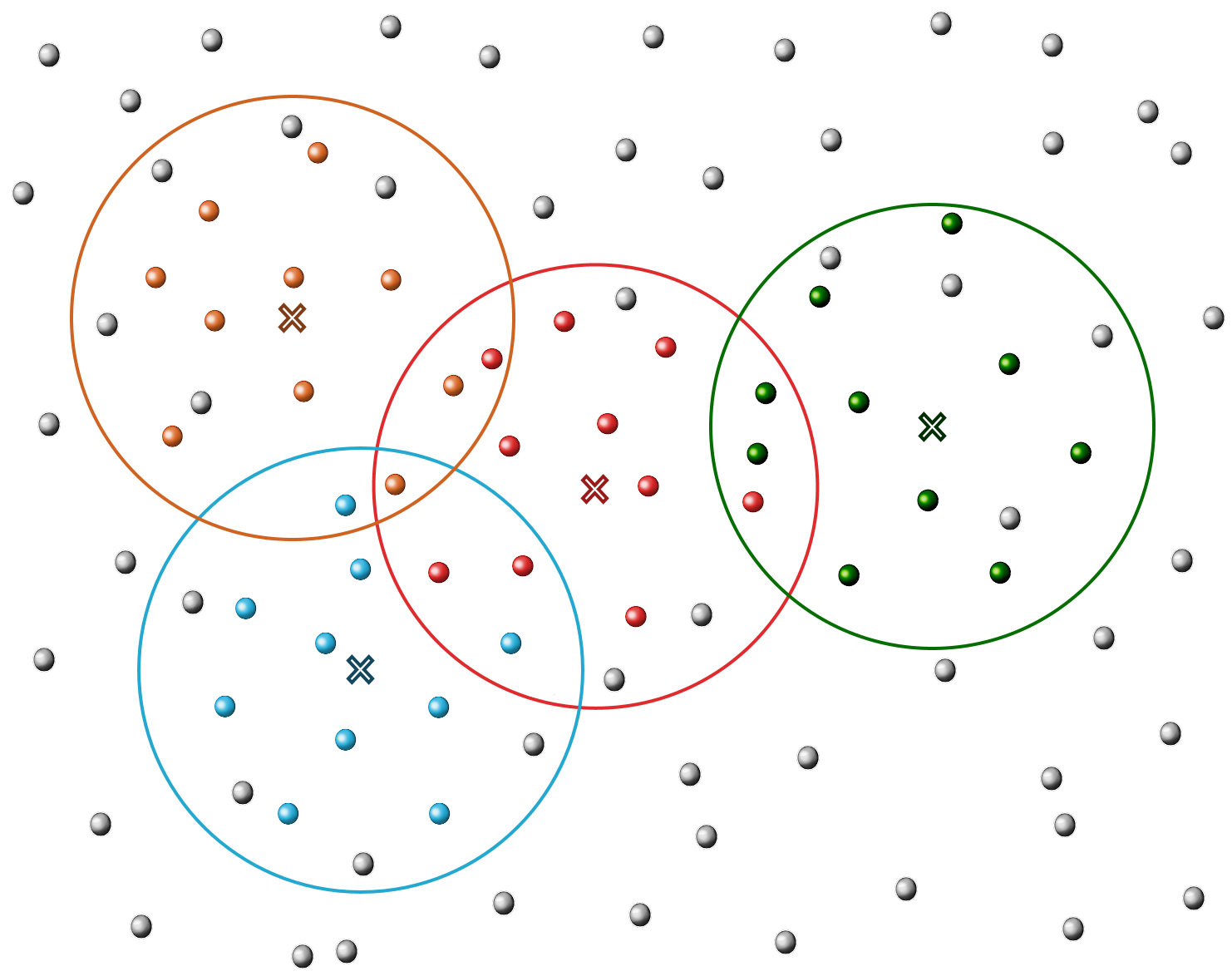}
	\caption{\cor{Illustration of mesoscopic particles, being subsets of $N_\mathrm{mic}$ microscopic particles within a radius of $R\simeq\sigma$. The colors of the microscopic particles indicate which mesoscopic particle they are associated with, with grey particles belonging to mesoscopic particles not marked in this figure. Each mesoscopic particle is characterized by its center-of-mass position, marked with a cross, its center-of-mass momentum and an enthalpy emerging from the random motion of the associated microscopic particles. To account for density fluctuations within the fluid, mesoscopic particles may overlap.}}
	\label{fig:illustration_of_mesoscopic_particles}
\end{figure}

Even though the essence of MPH is nothing but a mathematical trick it has a insightful physical interpretation which we want to highlight at first. One of the basic principles of hydrodynamics is a distinct hierarchy of scales. While the  equations describing an ideal gas are only valid on a macroscopic scale $L$ which is much larger than the mean free path of a single particle $\lambda$, it is possible to introduce a mesoscopic scale $\sigma$ with $\lambda \ll \sigma \ll L$. Considering a fluid element on that scale, it would inhabit a huge number of particles which would quickly establish thermodynamical equilibrium (according to local equilibrium hypothesis). This allows us to define thermodynamic quantities such as pressure and energy density. Due to this scale hierarchy we introduce a mesoscopic particle with a radius $R\simeq\sigma$ as an accumulation of $N_\text{mic}$ microscopic particles\cor{, as illustrated in \autoref{fig:illustration_of_mesoscopic_particles}}. The position of such a mesoscopic particle is defined by the center of mass of the assigned microscopic particles and its momentum is given \cor{by their} average velocity times their total mass $M$.
\par
On scales smaller than $\sigma$ the model will necessarily break down. But this actually does not pose any real restrictions to our model, since the hydrodynamical equations by themselves are only an approximation valid for appropriately large scales. \cor{Hence, one has to choose $\sigma$ sufficiently small such that the validity scale of the model lies below the hydrodynamical scale $L$. When calculating the density contrast power spectrum in \autoref{Tree-Level Power Spectrum} we will in fact take the limit $\sigma\rightarrow 0$ at the very end, corresponding to the hydrodynamic limit.}
\par
Since the position and momentum are obtained by averaging over the microscopic degrees of freedom we have to assign another property which characterizes the random microscopic nature. Information on these random distributions of microscopic are stored in the stress-energy tensor. In case of an isotropic gas the stress energy tensor is fully characterized by the pressure. For an ideal gas, the equation of state connecting pressure $P$ and internal energy density $\varepsilon$ is given by
\begin{equation}
	P = (\gamma - 1) \varepsilon,  \label{pressure equ}
\end{equation}
where $\gamma$ denotes the adiabatic index which can be expressed in terms of the number of degrees of freedom per particle as $\gamma = \frac{f+2}{f}$. The pressure and internal energy density can be combined into the enthalpy density 
\begin{equation}
	h = \varepsilon+P \label{enthalpy},
\end{equation}
and thus this allows us to preserve all necessary information about a fluid element by assigning the enthalpy $H$ as a third property to the description of a mesoscopic particle.
\par
To derive the equations of motions from the hydrodynamical equations, we need some smooth description for the mesoscopic particles. Therefore, we make use of the following trick: Imagine that the number density of the microscopic particles which we assign to the $j$-th mesoscopic particle is proportional to a smooth kernel function $W_{j}(\vec{r})$. A good candidate for such a kernel would be a Gaussian with the width chosen to be our mesoscopic scale $\sigma$. Like in our naive picture above we apply the same width to each particle, implying the kernel function to be of the form $W_{j}(\vec{r})=W(\vec{r}-\vec{r}_{j})$, where $\vec{r}_{j}$ denotes the position of the $j$-th mesoscopic particle. In the following discussion we assume a Gaussian kernel function, i.e.
\begin{equation}
	W(\vec{r})=\frac{1}{(2\pi\sigma^2)^{3/2}} \exp\bigg\{-\frac{\vec{r}^2}{2\sigma^2}\bigg\}.  \label{MPH: Kernel W(r)}
\end{equation}
Since we assign each microscopic particle to exactly one mesoscopic particle, we can approximate the total number density\footnote{In contrast to earlier papers, we denote the number density by $n$ and the mass density by $\rho$.} of microscopic particles as
\cor{\begin{align}
	n_{\mathrm{mic}}(\vec{r}) & \approx \sum_{i} N_\mathrm{mic} W_{i}(\vec{r})  = N_{\mathrm{mic}}  n_{\mathrm{mes}}(\vec r),   \label{density}
\end{align}}
where the sum over all mesoscopic particles is implied. Hence, the number density of microscopic particles is proportional to the number density of the mesoscopic particles. Similarly, any other density expression $a(\vec{r})$  can be decomposed into its single particle contributions $A_i$,
\begin{equation}
	a(\vec{r}) \approx \sum_{i} A_i W_i(\vec{r}).
\end{equation}

%%%%%%%%%%%%%%%%%%%%
\subsection{Equations of Motion for Mesoscopic Particles}
\label{EOM's Main}
Having an explicit picture of a mesoscopic particle in mind, we are now in the position to derive the equations of motion. Since these should fulfill the hydrodynamic equations on a macroscopic level, we consider the continuity, Euler and energy equations in the form
\begin{align}
	&\frac{\mathrm{d}}{\mathrm{d} t} \rho + \rho \nabla_r \cdot \vec{u} \label{MPH: ContinuityEq} = 0 , \\
	&\rho \frac{\mathrm{d}}{\mathrm{d} t} \vec{u} + \nabla_r P = -\rho\nabla_r\phi_\mathrm{g}	 \label{MPH: EulerEq},\\
	& \frac{\mathrm{d}}{\mathrm{d} t} \varepsilon + (\varepsilon + P) \nabla_r \cdot \vec{u} = 0  \label{MPH: EnergyEq}.
\end{align}
Here $\rho(\vec{r}, t)$ and $\vec{u}(\vec{r}, t)$ denote the mass density and the velocity of the cosmic fluid while $\phi_\mathrm{g}(\vec{r}, t)$ represents its gravitational potential. Note that KFT follows the motion of all individual mesoscopic particles. Hence, we can neglect the continuity equation since mass conservation will be fulfilled by construction. In the following, we exemplify the procedure of deriving the mesoscopic equations of motion for the momentum and enthalpy only for a single term and otherwise state the results while referring for more details to \autoref{EOM's Appendix}. 
\par
The momentum equation of motion for the $j$-th particle is obtained from the Euler equation \eqref{MPH: EulerEq}. For its derivation, we assume that we are in the rest frame of the corresponding particle. To obtain the contributions of the $j$-th particle we multiply the Euler equation with a projection function $\Pi_j(\vec{r})$ and integrate over the whole space. Such a projection function should be dimensionless and otherwise have similar characteristics as the kernel function $W(\vec{r})$ we introduced before. Thus, we choose the projection function as
\begin{equation}
	\Pi(\vec{r}) = 2^{3/2} \exp\bigg\{-\frac{\vec{r}^2}{2\sigma^2}\bigg\},   \label{MPH: Projection Pi(r)}
\end{equation}
where the prefactor is chosen such that  $\int_r\Pi(\vec{r})W(\vec{r})=1$.
\par
Performing the integration by using expression \eqref{density} for the number density, we get for the first term in \eqref{MPH: EulerEq}
\begin{align}
	\int \mathrm{d}^3r\ \Pi_j(\vec{r}) \rho(\vec{r}) \frac{\mathrm{d}}{\mathrm{d}t} \vec{u}(\vec{r}) &= \int \mathrm{d}^3r\ \Pi_j(\vec{r}) \sum_{i} M \dot{ \vec{u} }_i W_i(\vec{r}) \\
	& \approx M \dot{\vec{u}}_j  \int \mathrm{d}^3r\ \Pi_j(\vec{r}) W_j(\vec{r})   \nonumber\\
	& = \dot{\vec{p}}_{j} .  \nonumber
\end{align}
For the approximation in the second line to be valid the overlap between the mesoscopic particles\cor{, indicated in \autoref{fig:illustration_of_mesoscopic_particles},} must be sufficiently small which means the parameters of the model must be chosen carefully as we will discuss later.
\par
The other two terms can be treated analogously. In the end one finds an equation of motion of the form 
\begin{equation}
	\dot{p}_j = -M\nabla_r V(\vec{r})\big|_{r=r_j},   \label{mes momentum equation}
\end{equation}
where the interaction potential splits up into contributions from pressure and gravity,
\begin{equation}
	V(\vec{r}) \coloneqq V_\mathrm{p}(\vec{r}) + V_\mathrm{g} (\vec{r}),
\end{equation}
with
\begin{align}
	V_\mathrm{p}(\vec{r}) &= \sum_{i} \frac{\gamma-1}{\gamma} \frac{H_i}{M}  \int \mathrm{d}^3r'\ \Pi\big(\vec{r}'\big)  W\big(\vec{r}'-(\vec{r}_i-\vec{r})\big) ,\label{V_p}\\
	V_\mathrm{g}(\vec{r}) &= \sum_i GM \int \mathrm{d}^3 r' \mathrm{d}^3r''\ \frac{1}{|\vec{r}''-\vec{r}'|} \Pi\big(\vec{r}''\big) W(\vec{r}'')    W\big(\vec{r}'- (\vec{r}_i-\vec{r})\big)     \label{V_g}.
\end{align}
For this to hold we considered a Newtonian gravitational potential to describe the attractive force between the particles.
\par
The equation of motion for the enthalpy results from the energy equation and is found to be
\cor{\begin{equation}
	\dot{H}_j = - \nabla_{r_j} \sum_i (\gamma - 1)H_i (u_i-u_j) \int \mathrm{d}^3r'\ \Pi(\vec{r}') W\big( \vec{r}' - (\vec{r}_i-\vec{r}_j) \big) . \label{mes energy equation}
\end{equation}}
The equations of motion \eqref{mes momentum equation} and \eqref{mes energy equation} are exactly of the same form (disregarding the additional gravitational potential) as the findings in \cite{viermann_model_2018}.
\par
At this point we should uncover the limitation of the model. Since the microscopic distribution of the mesoscopic particles are described by a kernel function of finite width, they will overlap. Furthermore, the overlap will vary depending on the local (macroscopic) number density. But since we always use the same projection function $\Pi$ we will sometimes overestimate and sometimes underestimate the contributions to the $j$-th particle. For our calculations to be a good approximation we would need that the overlap is nearly the same everywhere, which means that the system is close to homogeneity. In \cite{viermann_model_2018} the same approximation appears when the inverse density is replaced by its mean value, i.e. $\frac{1}{n(\vec{r})} \approx \frac{1}{\bar{n}}$. Nevertheless, for the application to cosmology we assume that this is a reasonable approximation, due to the homogeneity on large scales.

%%%%%%%%%%%%%%%%%%%%
\subsection{Special Case: The Isothermal Fluid}
\label{subsection isothermal fluid}
To simplify the treatment of the mesoscopic equations of motion in KFT,
let us consider the special case of an isothermal \cor{fluid, which} implies that the equation of state takes the form
\begin{equation}
	P=c_\mathrm{s}^2\rho,
	\label{isothermal_equation_of_state}
\end{equation}
where $c_\mathrm{s}$ denotes the isothermal speed of sound. Using \eqref{pressure equ} and \eqref{enthalpy}, this implies a constant enthalpy per mesoscopic particle,
\begin{equation}
	H_i = \frac{\gamma}{(\gamma-1)} m c_\mathrm{s}^2.  \label{H_i equ}
\end{equation}
This effectively reduces the degrees of freedom of a mesoscopic particle to its position and momentum, allowing us to treat them formally very similarly to microscopic particles within KFT. Note that while an isothermal description is only a crude approximation to the full gas dynamics of baryonic matter, it should still capture the general qualitative behaviour of matter that is subject to pressure. In a future paper we will show how this treatment can be extended to general ideal gas dynamics.
\par
The isothermal model is now fully described by the free motion of the mesoscopic particles plus perturbations to the particle momenta obtained from \eqref{mes momentum equation} using \eqref{H_i equ} in the pressure potential $V_\mathrm{p}$. Defining the one-particle \cor{potentials $v_\mathrm{p}$ and $v_\mathrm{g}$ by
\begin{equation}
	V_\mathrm{p}(\vec{r}) \eqqcolon \sum_i v_\mathrm{p}(\vec{r}-\vec{r}_i), \qquad V_\mathrm{g}(\vec{r}) \eqqcolon \sum_i v_\mathrm{g}(\vec{r}-\vec{r}_i),
\end{equation}
we can find explicit expression for them} by evaluating the integrals appearing in \eqref{V_p} and \eqref{V_g}, and inserting explicit functions for $W(\vec{r})$ and $\Pi(\vec{r})$ as given by \eqref{MPH: Kernel W(r)} and \eqref{MPH: Projection Pi(r)}. 
\par
For our later purposes, we are particularly interested in the Fourier transformed one-particle potentials. Using the convention
\begin{equation}
	f(\vec{k}) = \int \mathrm{d}^3r\  \mathrm{e}^{\mathrm{i}\vec{k}\cdot\vec{r}} f(\vec{r}),
\end{equation}
they are found to take the form 
\begin{align}
	v_\mathrm{p}(\vec{k}) &= C_\mathrm{p} \exp\bigg\{-\sigma^2\vec{k}^2\bigg\},    \\
	v_\mathrm{g}(\vec{k}) &=  - \frac{C_\mathrm{g}}{\vec{k}^2+\epsilon^2}\exp\bigg\{-\frac{3}{4}\sigma^2\vec{k}^2\bigg\}
\end{align}
with the constant prefactors given by
\begin{align}
	C_\mathrm{p}&=(4\pi)^{3/2} c_\mathrm{s}^2\sigma^3 \label{C_p}, \\
	C_\mathrm{g}&= 4\pi GM.  \label{C_g}
\end{align}
While the integrations leading to the pressure potential are straight forward, the calculations for the gravitational potential can be found in \autoref{Potentials Appendix}. The parameter $\epsilon$ describes an IR cutoff which has to be taken to zero at the very end of any calculation. However, as long as $k\neq0$, $\epsilon$ can immediately be set to zero. For small values of $|\vec{k}|$ (corresponding to large distances with $|\vec{R}|\gg\sigma$) the exponential can be approximated by unity and the result for the gravitational potential becomes identical to the potential of a point particle, as one would expect.

%%%%%%%%%%
\subsubsection*{Jeans Length}
Reflecting the form of the two potentials, we notice that they depend on the model specific quantities $\sigma$ and $N_\mathrm{mic}$ (note that $M$ carries some intrinsic dependence on $N_\mathrm{mic}$) which are related to each other. To visualize this point, let us assume a system of average mesoscopic number density $\bar{n}$ and say we set the mesoscopic scale $\sigma$ to some specific value. What we may ask is if we are free to choose the microscopic particle number $N_\mathrm{mic}$. The answer is no: If $N_\mathrm{mic}$ is too big, the mesoscopic particle will not cover enough space to inhabit all $N_\mathrm{mic}$ microscopic particles. On the other side, if we select it too small, the mesoscopic particles will strongly overlap such that the approximations we made for the derivation of the equations of motion are no longer valid. Thus, we have to deduce how to choose the microscopic number $N_\mathrm{mic}$ depending on the underlying system. A natural way to do this is to compare it with characteristic scales of the system. A good candidate in our case is the Jeans length $\lambda_\mathrm{J}=\frac{2\pi}{k_\mathrm{J}}$ with
\begin{equation}
	k_\mathrm{J} = \frac{\sqrt{4\pi G \bar{\rho}}}{c_\mathrm{s}} \label{exact JL}
\end{equation}
that describes the scale at which gravity and pressure effects equalize each other \cite{bartelmann_theoretical_2013}.
\par
For comparison it is reasonable to rewrite $k_\mathrm{J}$ in terms of the potential prefactors $C_\mathrm{p}$ and $C_\mathrm{g}$. The number $N_\mathrm{mic}$ should be proportional to $\sigma^3$, which gives some volume measure for the space that is covered by a mesoscopic particle, as well as the average number density $\bar{n}_\mathrm{mic}$ of microscopic particles. Thus
\begin{equation}
	N_\mathrm{mic}=\alpha\sigma^3 \bar{n}_\mathrm{mic} \label{Jeans coefficient}
\end{equation}
with some constant $\alpha$. Inserting this into $C_\mathrm{p}$, we obtain
\begin{equation}
	\frac{C_\mathrm{g}}{C_\mathrm{p}}  =  \frac{\alpha}{(4\pi)^{3/2}} \frac{4\pi G\bar{\rho}}{ c_\mathrm{s}^2} = \frac{\alpha }{(4\pi)^{3/2}} k_\mathrm{J}^2,   \label{JeansLength of X}
\end{equation}
where we used $\bar{\rho}= M\bar{n} = \frac{M\bar{n}_\mathrm{mic}}{N_\mathrm{mic}}$. For a good choice of our model parameters we should be able to recover the Jeans length. That is, the Jeans length should correspond to the scale at which the pressure and gravitational potential of the model are equal, i.e. $k_\mathrm{J}=k^{\mathrm{MPH}}_\mathrm{J}$ with $k^{\mathrm{MPH}}_\mathrm{J}$ given by the solution of 
\begin{align}
	&\ v_\mathrm{p}(k) + v_\mathrm{g}(k) = 0 \label{MPH Jeans length}\\
	\Leftrightarrow&\  (k^{\mathrm{MPH}}_\mathrm{J})^2 = \frac{\alpha}{(4\pi)^{3/2}} k_\mathrm{J}^{2} \exp\bigg\{- \bigg(\frac{\sigma k^\mathrm{MPH}_\mathrm{J}}{2}\bigg)^2 \bigg\} . \nonumber
\end{align}
Demanding the equality of the two notions of the Jeans length, we find
\begin{equation}
	\alpha = (4\pi)^{3/2} \exp\bigg\{ \bigg(\frac{\sigma k_\mathrm{J}}{2}\bigg)^2 \bigg\}. \label{Jeans alpha}
\end{equation}
From this we infer that
\begin{equation}
	\alpha = (4\pi)^{3/2} \qquad \mathrm{while} \qquad \sigma \ll k_\mathrm{J}^{-1},  \label{alpha}
\end{equation}
which we demand according to our discussion of the scale hierarchy.

%%%%%%%%%%
\subsubsection*{Generalisation to an Expanding Space-Time}
We are particularly interested in an application to cosmology, thus we have to take into account the expanding nature of the universe. To include this property, we have to investigate how the expansion modifies the interaction potential of the mesoscopic particles. For this purpose, we should first think about how this affects the idea of a mesoscopic particle. Since the expansion forces microscopic particles to move apart, the radius of a mesoscopic particle should become time-dependent and increase with the rate of expansion described by the scale factor $a(t)$. Hence, we infer $\sigma(t)=a(t)\sigma_0$. 
\par
For the treatment of an expanding space it is convenient to work in comoving coordinates $\vec{q}$, which are related to physical coordinates by $\vec{r}=a\vec{q}$. Furthermore we use the dimensionless time coordinate $\eta=\ln a$. Choosing $a_\mathrm{i}=1$ for some initial time $t_\mathrm{i}$, it follows that $\eta_\mathrm{i}=0$. Then, the kernel and projection functions become
\cor{\begin{align}
	W(\vec{q},\eta) &=   \frac{1}{a(\eta)^3} \frac{1}{(2\pi\sigma_0^2)^{3/2}}   \exp\bigg\{-\frac{\vec{q}^2}{2\sigma_0^2}\bigg\},   \\
\Pi(\vec{q}) &=  2^{3/2}  \exp\bigg\{-\frac{\vec{q}^2}{2\sigma_0^2}\bigg\}.
\end{align}}
Inserting these results into the expressions \eqref{V_p} and \eqref{V_g} for the interaction potentials, and noting that $\mathrm{d}^3r=a^3\mathrm{d}^3q$, we see that the structure of the pressure potential does not change. On the other hand, the gravitational potential gets rescaled by a factor of $\frac{1}{a}$ due to the $\frac{1}{|\vec{r}-\vec{r}'|}$-term. Thus, we find
\cor{\begin{align}
	v_\mathrm{p}(\vec{k}) &= C_{\mathrm{p}}  \exp\bigg\{-\sigma_0^2\vec{k}^2\bigg\} ,	\label{pressure potential in expanding space-time}	\\
	v_\mathrm{g}(\vec{k},\eta) &= -  \frac{1}{a(\eta)} \frac{C_{\mathrm{g}}}{\vec{k}^2+\epsilon^2}\exp\bigg\{-\frac{3}{4}\sigma_0^2\vec{k}^2\bigg\} , \label{gravitational potential in expanding space-time}
\end{align}}
where from now on $k$ denotes the Fourier conjugate to the comoving coordinate $q$.
\par
\cor{The corresponding Hamiltonian of a mesoscopic particle in an expanding space-time is found by adopting the derivation in \cite{bartelmann_trajectories_2015} to our choice of time coordinate and the additional pressure potential. It reads
\begin{equation}
	\mathcal{H} = \frac{\vec{p}^2}{2g(\eta)} + V_{\mathrm{eff}}(\vec{q},\eta)	\label{MPH: Hamiltonian}
\end{equation}
with the canonically conjugated momentum $\vec{p}=g(\eta)\frac{\mathrm{d}\vec{q}}{\mathrm{d}\eta}$ and
\begin{equation}
	g = a^2 \frac{H}{H_\mathrm{i}} ,
\end{equation}
where $H$ denotes the Hubble function and $H_\mathrm{i} \coloneqq H(\eta_\mathrm{i})$.} The effective potential is defined as
\begin{equation}
	V_{\mathrm{eff}}(\vec{q},\eta)\coloneqq\frac{a(\eta)^2}{H_\mathrm{i}^2g(\eta)} \Big( V_\mathrm{p}(a\vec{q}) + V_\mathrm{g}(a\vec{q}) \Big),  \label{eff Potential}
\end{equation}
\cor{suggesting to rewrite the potential prefactors $C_{\mathrm{p}}$ and $C_{\mathrm{g}}$, defined in \eqref{C_p} and \eqref{C_g}, as}
\begin{align}
	C_{\mathrm{p},\mathrm{eff}} &\coloneqq \frac{C_\mathrm{p}}{H_\mathrm{i}^2} = \frac{c_\mathrm{s}^2}{H_\mathrm{i}^2\bar{n}}  \label{effective Cp},\\
	C_{\mathrm{g},\mathrm{eff}} &\coloneqq \frac{C_\mathrm{g}}{H_\mathrm{i}^2} = \frac{3}{2\bar{n}}.   \label{effective Cg}
\end{align}
\cor{For the calculation of $C_\mathrm{p,eff}$ we inserted $\sigma^3=\frac{1}{(4\pi)^{3/2}\bar{n}}$, which is obtained by combining equations \eqref{Jeans coefficient}, \eqref{alpha} and \eqref{density}. For $C_\mathrm{g,eff}$ we used that
the initial value of the dimensionless matter density parameter is well-approximated by unity, $\Omega_\mathrm{m,i}=\frac{8\pi GM}{3 H_\mathrm{i}^2}\bar{n}\approx1$, since we fix the initial time to an instant in the matter-dominated phase. The only model-dependent quantity in \eqref{effective Cp} and \eqref{effective Cg}} is the mesoscopic number density. By direct computation one finds that this drops out for an explicit expression of the power spectrum of pure baryonic matter. If one considers the more general case of mixed matter, ratios of number densities will affect the power spectrum, which cancels the dependence on the choice of $N_\mathrm{mic}$ as well, leading to a consistent model-independent picture.
\par
From the free Hamiltonian $\mathcal{H}_0=\frac{\vec{p}^2}{2g(\eta)}$ one can derive the retarded Green's function of the free equation of motion to be 
\begin{equation}
	\mathcal{G}(\eta,\eta') = 
	\begin{pmatrix}
		\mathbbm{1}_3 & g_{qp}(\eta,\eta')\\
		0 & \mathbbm{1}_3 \\
	\end{pmatrix}
	\theta(\eta-\eta'), \label{MPH: propagator}
\end{equation}
where
\begin{equation}
	g_{qp}(\eta,\eta') = \int_{\eta'}^{\eta} \mathrm{d}\tilde{\eta}\  \frac{1}{g(\tilde{\eta})}.   \label{MPH: prop g_qp}
\end{equation}
In an EdS universe the relative expansion rate is given by $H=H_\mathrm{i}a^{-3/2}$, leading to
\begin{equation}
	g(\eta)=a(\eta)^{1/2}=\mathrm{e}^{\eta/2}.
	\label{EdS g factor}
\end{equation}
Inserting this into the expression for $g_{qp}$, we end up with
\begin{equation}
	g_{qp}(\eta,\eta')= -2\big( \mathrm{e}^{-\eta/2}-\mathrm{e}^{-\eta'/2} \big).
	\label{qp-component of retarded Green's function}
\end{equation}
\par
\cor{By using the relation \eqref{EdS g factor} in \eqref{eff Potential}, we can further infer explicit expressions for the Fourier transformed effective one-particle potentials in an EdS cosmology,
\begin{align}
	v_{\mathrm{p},\mathrm{eff}}(\vec{k},\eta) &= C_{\mathrm{p},\mathrm{eff}} \, \mathrm{e}^{3\eta/2} \exp\bigg\{-\sigma_0^2\vec{k}^2\bigg\} ,
	\label{effective pressure potential in EdS}	\\
	v_{\mathrm{g},\mathrm{eff}}(\vec{k},\eta) &= - \frac{C_{\mathrm{g},\mathrm{eff}}}{\vec{k}^2+\epsilon^2} \, \mathrm{e}^{\eta/2} \exp\bigg\{-\frac{3}{4}\sigma_0^2\vec{k}^2\bigg\} .
	\label{effective gravitational potential in EdS}
\end{align}
Note that the different relative time-dependencies of $v_{\mathrm{p},\mathrm{eff}}$ and $v_{\mathrm{g},\mathrm{eff}}$ render the Jeans length in an expanding space-time time-dependent. The corresponding wavenumber $k_{\mathrm{J},\mathrm{eff}}$ is obtained by solving $v_{\mathrm{p},\mathrm{eff}} + v_{\mathrm{p},\mathrm{eff}} = 0$ for $\epsilon = 0$ and $\sigma_0 \ll k_{\mathrm{J},\mathrm{eff}}^{-1}$. This yields
\begin{equation}
	k_{\mathrm{J},\mathrm{eff}}(\eta) = \mathrm{e}^{-\eta/2} \, \sqrt{\frac{C_{\mathrm{g},\mathrm{eff}}}{C_{\mathrm{p},\mathrm{eff}}}} = \mathrm{e}^{-\eta/2} \, \sqrt{\frac{3}{2}} \, \frac{H_\mathrm{i}}{c_\mathrm{s}} ,
	\label{effective Jeans wavenumber}
\end{equation}
where we used \eqref{effective Cp} and \eqref{effective Cg} in the second step.}

%%%%%%%%%%%%%%%%%%%%%%%%%%%%%%%%%%%%%%%%
\section{Application to KFT}
\label{Section KFT}

%%%%%%%%%%%%%%%%%%%%
\subsection{\cor{General Framework of KFT}}
The standard description of cosmological structure formation is based on a macroscopic theory of hydrodynamics. In contrast, the KFT approach tackles this issue on a microscopic level. Therefore, a generating functional is constructed which encapsulates both the dynamics and the initial statistics of a non-equilibrium many-particle \cor{system. In} the following, we summarize the conceptually most important aspects of the KFT formalism and refer for a more detailed description to \cite{bartelmann_microscopic_2016,bartelmann_kinetic_2017,bartelmann_analytic_2017}.

%%%%%%%%%%
\subsubsection*{The Generating Functional}
We consider a system of N particles confined to a volume $V$. The individual particles are described by their phase-space coordinates $\vec{x}_j\coloneqq(\vec{q}_j,\vec{p}_j)$, labeled by $j=1,\dotsc,N$, with $\vec{q}_j$ denoting the comoving position and $\vec{p}_j$ the canonically conjugated momentum of the $j$-th particle. It is conventional to bundle the phase-space coordinates into a phase-space tensor
\begin{equation}
	\boldsymbol{x} \coloneqq \vec{x}_j \otimes \vec{e}_j
\end{equation}
with the canonical base $\{\vec{e}_j\}$ in $N$ dimensions and entries $(e_j)_i=\delta_{ij}$. Furthermore, summation over repeated indices is implied \cor{if not explicitly stated otherwise}. This notation can also be applied to other quantities. Throughout this paper, bold-faced symbols always denote tensors combining the contributions from all N particles as defined above. A scalar product between two such tensors can be defined as
\begin{align}
	\boldsymbol{a} \cdot \boldsymbol{b}  = \vec{a}_j \cdot \vec{b}_j.
\end{align}
\par
Now, we can formulate a generating functional by integrating over all possible trajectories, where a Dirac delta distribution ensures the classical equation of motion $\boldsymbol{E}[\boldsymbol{x}]=0$, given by Hamilton's equation with the Hamiltonian given in \eqref{MPH: Hamiltonian}, to hold. In addition, stochasticity enters the expressions by averaging over the initial conditions \cor{at time $t_\mathrm{i}$} according to an initial phase-space probability $\mathcal{P}(\boldsymbol{x}^{(\mathrm{i})})$. The generating functional reads
\cor{\begin{equation}
	Z[\boldsymbol{J},\boldsymbol{K}] = \int \! \mathrm{d}\boldsymbol{x}^{(\mathrm{i})} \ \mathcal{P}(\boldsymbol{x}^{(\mathrm{i})}) \int_{\boldsymbol{x}^{(\mathrm{i})} } \!\!\!\! \mathcal{D}\boldsymbol{x}(t) \int \! \mathcal{D}\boldsymbol{\chi}(t) \, \exp \bigg\{ \mathrm{i} S[\boldsymbol{x},\boldsymbol{\chi}] + \mathrm{i} \int_{t_\mathrm{i}}^\infty \!\!\!\! \mathrm{d}t \, \big(\boldsymbol{\chi} \cdot \boldsymbol{K} + \boldsymbol{x} \cdot \boldsymbol{J} \big) \bigg\},
	\label{Generating functional of microscopic correlators}
\end{equation}
where an auxiliary field $\boldsymbol{\chi}$ with components $\vec{\chi}_j = \bigl(\vec{\chi}_{q_j}, \vec{\chi}_{p_j}\bigr)$ was introduced to express the delta distribution as a path integral with respect to the action
\begin{equation}
	S[\boldsymbol{x},\boldsymbol{\chi}] \coloneqq \int_{t_\mathrm{i}}^\infty \mathrm{d}t \,\boldsymbol{\chi}(t) \cdot \boldsymbol{E}[\boldsymbol{x}(t)] .	
\end{equation}}
Furthermore, we introduced the source fields $\boldsymbol{K}(t)$ and $\boldsymbol{J}(t)$ for the auxiliary field $\boldsymbol{\chi}(t)$ and the \cor{phase-space} trajectory $\boldsymbol{x}(t)$. From this expression the field correlators can be derived by taking functional derivatives with respect to the respective source fields and setting all source fields to zero in the end,
\cor{\begin{align}
 \bigl\langle \boldsymbol{x}(t) \otimes \dotsm \boldsymbol{\chi}(t') \otimes \dotsm \bigr\rangle 
 = \frac{\delta}{\mathrm{i}\delta \boldsymbol{J}(t)} \otimes \dotsm \frac{\delta}{\mathrm{i}\delta \boldsymbol{K}(t')} \otimes \dotsm Z[\boldsymbol{J},\boldsymbol{K}] \bigg |_{\boldsymbol{J},\boldsymbol{K}=0} .
\end{align}}
Likewise, as in quantum field theory, the functional $W[\boldsymbol{J},\boldsymbol{K}]=\ln Z[\boldsymbol{J},\boldsymbol{K}] $ is the generating functional for cumulants, i.e. connected correlators.

%%%%%%%%%%
\subsubsection*{The Initial Distribution}
In view of cosmological structure formation the initial distribution of the microscopic particles has to be chosen in such a way that it reflects the homogeneous and isotropic nature of the universe on large scales. Assuming the deviations of the initial density and momentum fields are Gaussian distributed, we can map this macroscopic behavior on the microscopic degrees of freedom by using a Poisson sampling. This implies spatial and momentum correlations as well as cross-correlations between the single particle coordinates.
\par
\cor{The derivation is shown explicitly in \cite{bartelmann_microscopic_2016} and yields}
\begin{equation}
	\mathcal{P}(\boldsymbol{x}^{(\mathrm{i})}) \coloneqq \frac{V^{-N}}{\sqrt{(2\pi)^{3N} \det C_{pp}}} \mathcal{C} (\boldsymbol{p}) \exp \bigg\{ -\frac{1}{2} \boldsymbol{p}^\intercal C^{-1}_{pp} \boldsymbol{p} \bigg\},     \label{Initial distr}
\end{equation}
\cor{where} $C_{pp}$ is the covariance matrix of the initial momentum-momentum correlations, and the factor $\mathcal{C}(\boldsymbol{p})$ is given by a polynomial expression depending on the initial density-density and density-momentum correlations.
\par
\cor{If the initial density and momentum fields are linearly related, which is an excellent approximation for the initial conditions of cosmic structure formation, then all three types of auto- and cross-correlations can be expressed in terms of the initial density contrast power spectrum $P_\delta^{(\mathrm{i})} \coloneqq P_\delta(k,t_\mathrm{i})$, where $\delta\coloneqq\frac{n-\bar{n}}{\bar{n}}$ and}
\begin{equation}
	 \langle \delta(\vec{k}) \delta(\vec{k}') \rangle = (2\pi)^3 \delta_\mathrm{D} (\vec{k}+\vec{k}') P_\delta(k,t) .  \label{Power Spectrum}
\end{equation}
\cor{Here, $\bar{n}$ denotes the mean number density.}

%%%%%%%%%%
\subsubsection*{Collective Fields}
Since we are actually not interested in the statistics of microscopic fields but rather in macroscopic properties of the ensemble, it is advisable to introduce collective fields. \cor{The minimum required set of collective fields consists of the number density $\Phi_n$, which carries the information on the particles' positions, and the response field $\Phi_B$, which encodes how the particle momenta are changed by a given interaction potential.} In Fourier space they take the form
\cor{\begin{align}
	\Phi_n(1) =& \sum_{j=1}^{N} \mathrm{e}^{\mathrm{i}\vec{k}_1\cdot\vec{q}_j(t_1)}  ,\\
	\Phi_B(1) =& \sum_{j=1}^{N} \mathrm{i}\vec{k}_1 \cdot \vec{\chi}_{p_j}(t_1) \; \mathrm{e}^{-\mathrm{i}\vec{k}_1\cdot\vec{q}_j(t_1)} ,
\end{align}
where we introduced the abbreviation $(\pm l) \coloneqq (\pm\vec{k}_l,t_l)$ for the Fourier space and time arguments of a function.}
\par
\cor{To extract statistical information on these collective fields from the generating functional, we define the respective collective-field operators $\hat{\Phi}_n$ and $\hat{\Phi}_B$ by replacing all occurrences of the fields $\boldsymbol{x}$ and $\boldsymbol{\chi}$ by functional derivatives with respect to their associated source fields $\boldsymbol{J}$ and $\boldsymbol{K}$,
\begin{align}
	\hat{\Phi}_n(1) \coloneqq& \sum_{j=1}^{N} \exp\biggl\{\mathrm{i}\vec{k}_1 \cdot \frac{\delta}{\mathrm{i} \delta \vec{J}_{q_j}(t_1)}\biggr\}  ,
	\label{collective number density field operator} \\
	\hat{\Phi}_B(1) \coloneqq& \sum_{j=1}^{N} \mathrm{i}\vec{k}_1 \cdot \frac{\delta}{\mathrm{i} \delta \vec{K}_{p_j}(t_1)} \, \exp\biggl\{\mathrm{i}\vec{k}_1 \cdot \frac{\delta}{\mathrm{i} \delta \vec{J}_{q_j}(t_1)}\biggr\} .
	\label{collective response field operator}
\end{align}
Acting with these operators on $W[\boldsymbol{J},\boldsymbol{K}] = \ln Z[\boldsymbol{J},\boldsymbol{K}]$ and setting the source fields to zero afterwards yields the collective-field cumulants,
\begin{align}
	G_{n \dotsm n B \dotsm B}(1, \dotsc, l_n, 1', \dotsc, l'_B) &\coloneqq \bigl\langle\Phi_n(1) \dotsm \Phi_n(l_n) \, \Phi_B(1') \dotsm \Phi_B(l'_B)\bigr\rangle_\textsc{c} 	\label{collective-field cumulants} \\
	&\,= \left. \prod_{u=1}^{l_n} \hat{\Phi}_n(u) \, \prod_{r=1}^{l_B} \hat{\Phi}_B(r') \, W[\boldsymbol{J}, \boldsymbol{K}] \, \right|_{\boldsymbol{J}, \boldsymbol{K}=0} \,. \nonumber
\end{align}
They describe the connected $l_n$-point correlations of the number density as well as their response to perturbations of the system at times $t'_1, \dotsc, t'_{l_B}$ caused by a given interaction potential.}
\par
\cor{The collective fields can further be used to rewrite the generating functional \eqref{Generating functional of microscopic correlators} in a way that facilitates a perturbative treatment of the particle interactions. For this, we first split the action into a free and an interacting part, $S = S_0 + S_\mathrm{I}$, where $S_0$ contains the free Hamiltonian equations of motion and $S_\mathrm{I}$ the contributions caused by interactions. The latter can then be expressed solely in terms of the two collective fields and the one-particle interaction potential $v$,
\begin{equation}
	S_\mathrm{I}[\boldsymbol{x},\boldsymbol{\chi}] 
	= - \int \mathrm{d}1 \, \Phi_n(1) \, v(-1) \, \Phi_B(-1)
	\label{interacting part of the action} ,
\end{equation}
using the short-hand notation
\begin{equation}
	\int \mathrm{d}l \coloneqq \int \frac{\mathrm{d}^3k_l}{(2\pi)^3} \, \int_{t_\mathrm{i}}^\infty \!\!\! \mathrm{d}t_l
\end{equation}
for integrals over Fourier space and time. Essentially, \eqref{interacting part of the action} describes the response of the system to the total interaction potential generated by the interactions between all the individual particles. In our case, $v$ is given by the sum of the effective one-particle potentials \eqref{effective pressure potential in EdS} and \eqref{effective gravitational potential in EdS}, $v = v_{\mathrm{p},\mathrm{eff}} + v_{\mathrm{g},\mathrm{eff}}$.}
\par
\cor{Using the collective-field operators \eqref{collective number density field operator} and \eqref{collective response field operator}, the interacting part of the action can now be replaced by the operator expression
\begin{equation}
	\hat{S}_\mathrm{I} \coloneqq - \int \mathrm{d}1 \, \hat{\Phi}_n(1) \, v(-1) \, \hat{\Phi}_B(-1) ,
\end{equation}
allowing us to pull it in front of the path integral in \eqref{Generating functional of microscopic correlators}. The generating functional can thus be written as
\begin{equation}
	Z[\boldsymbol{J}, \boldsymbol{K}] = \mathrm{e}^{\mathrm{i}\hat{S}_\mathrm{I}} Z_0[\boldsymbol{J}, \boldsymbol{K}], \label{Basics: Full Gen. Func. for Class. Fields}
\end{equation}
where the interaction operator $\hat{S}_\mathrm{I}$ is acting on the remaining free generating functional}
\begin{equation}
	Z_0[\boldsymbol{J}, \boldsymbol{K}] = \int \mathrm{d}\boldsymbol{x}^{(\mathrm{i})}\ \mathcal{P}(\boldsymbol{x}^{(\mathrm{i})}) \mathrm{e}^{\mathrm{i}\int \mathrm{d}t\ \boldsymbol{J} \cdot \boldsymbol{\bar{x}} }  . \label{Free Generating Functional}
\end{equation}
$\boldsymbol{\bar{x}}(t)$ is the solution to the linear free equation of motion and takes the form
\begin{equation}
	\boldsymbol{\bar{x}}(t) = \boldsymbol{\mathcal{G}}(t,t_\mathrm{i}) \boldsymbol{x}^{(\mathrm{i})} - \int_{t_\mathrm{i}}^{\infty} \mathrm{d}t'\ \boldsymbol{\mathcal{G}}(t,t') \boldsymbol{K}(t')
\end{equation}
with the $N$-particle retarded microscopic propagator
\begin{equation}
	\boldsymbol{\mathcal{G}}(t,t') = \mathcal{G}(t,t') \otimes \mathbbm{1}_N .
\end{equation}
Here $\mathbbm{1}_N$ denotes the $N$-dimensional unit matrix and $\mathcal{G}$ is given by \eqref{MPH: propagator}.
\par
\cor{In analogy to \eqref{collective-field cumulants}, acting with the collective-field operators on $W_0[\boldsymbol{J},\boldsymbol{K}] = \ln Z_0[\boldsymbol{J},\boldsymbol{K}]$ and setting the source fields to zero afterwards generates the collective-field cumulants of the freely evolving system, which we denote by $G^{(0)}_{n \dotsm n B \dotsm B}$. They can be calculated using a systematic expansion of the initial probability distribution \eqref{Initial distr} into contributions from different numbers of correlated particles. A diagrammatic approach to this is discussed in detail in \cite{fabis_kinetic_2018}.}
\par
\cor{The standard microscopic perturbative approach to KFT, detailed in \cite{bartelmann_microscopic_2016}, would now proceed by expanding the exponential in \eqref{Basics: Full Gen. Func. for Class. Fields} in orders in the interaction operator $\hat{S}_\mathrm{I}$. Although this approach has been used successfully to describe the structure formation in collisionless dark matter, see e.g. \cite{bartelmann_microscopic_2016,bartelmann_kinetic_2017}, its application to MPH in \cite{viermann_model_2018} indicated that any expansion to finite order in $\hat{S}_\mathrm{I}$ is probably insufficient to treat fluid dynamics consistently. Describing the structure formation in baryonic matter hence requires a different approach.}

%%%%%%%%%%%%%%%%%%%%
\subsection{\cor{Resummed KFT}}
\label{Section Resummation}
Recently, Lilow et al. \cite{lilow_resummed_2019} proposed an alternative approach dubbed Resummed KFT (RKFT) based on a reformulation of the KFT path integral which formally integrates out all microscopic degrees of freedom, leading to a theory formulated in terms of macroscopic quantities only. \cor{As a result}, even the lowest-order perturbative calculation within RKFT resums an infinite subset of terms appearing in the Taylor expansion of \cor{$e^{\mathrm{i} \hat{S}_\mathrm{I}}$}.
\par
\cor{A convenient consequence of approximating baryonic matter as an \emph{isothermal} fluid is that the mesoscopic particles obey Hamiltonian dynamics in this case -- just like dark matter particles. The only difference is that MPH imposes an additional pressure contribution to the interaction potential. For more general fluids, however, one would have to include the enthalpy (or equivalently the internal energy) as an additional degree of freedom of a mesoscopic particle. In that case, the particles would no longer obey Hamiltonian dynamics.}
\par
\cor{In this paper we restrict ourselves to isothermal fluids, while a discussion of the more general case is subject of future work. Crucially, the derivation of the RKFT formalism, as laid out in detail in \cite{lilow_resummed_2019}, thus remains completely unchanged. In the following, we will summarize this derivation with the minor simplification} that we choose the number density $n$ as the central quantity instead of the Klimontovich phase-space density $f$. This yields the same results as in \cite{lilow_resummed_2019} when computing cumulants of $n$, like the density contrast power spectrum we are interested in. \cor{Using the phase-space density would just allow us to also} calculate cumulants involving the momentum density and higher-order momentum moments.

%%%%%%%%%%
\subsubsection*{\cor{The Macroscopic Generating Functional}}
\cor{In RKFT, the generating functional \eqref{Generating functional of microscopic correlators} is reformulated as a path integral over the macroscopic number density field $n$ and a macroscopic auxiliary field $\beta$ rather than the microscopic fields $\boldsymbol{x}$ and $\boldsymbol{\chi}$. This is achieved by exploiting that the free evolution of the system is exactly solvable and that the interacting part of the action \eqref{interacting part of the action} depends on the microscopic fields only implicitly via the collective fields $\Phi_n$ and $\Phi_B$. Crucially, this reformulation is exact, and thus all information on the dynamics of the microscopic (or mesoscopic) particles is preserved.}
\par
\cor{Combining $n$ and $\beta$ into the combined macroscopic field $\phi \coloneqq (n,\beta)$ and introducing an associated macroscopic source field $M \coloneqq (M_n,M_\beta)$ instead of the microscopic source fields $\boldsymbol{J}$ and $\boldsymbol{K}$, the generating functional of macroscopic-field correlators takes the form
\begin{equation}
	Z_\phi[M]=\int \mathcal{D}\phi \, \exp\biggl\{ \mathrm{i} S_\Delta[\phi] + \mathrm{i} S_\mathcal{V}[\phi] + \int \mathrm{d}1 \, M^\top\!(1) \; \phi(-1) \biggr\} ,
	\label{RKFT Z}
\end{equation}
where $S_\Delta$ and $S_\mathcal{V}$ denote the propagator and vertex parts of the macroscopic action, respectively,
\begin{align}
	\mathrm{i} S_\Delta[\phi] &\coloneqq -\frac{1}{2} \int \mathrm{d}1 \int \mathrm{d}2 \; \phi^\top\!(-1) \; \Delta^{-1}(1,2) \; \phi(-2) , \\
	\mathrm{i} S_\mathcal{V}[\phi] &\coloneqq \sum_{\substack{l_\beta,l_n=0 \\ l_\beta+l_n\neq 2} }^{\infty} \frac{1}{l_\beta! \, l_n! } \prod_{u=1}^{l_\beta} \left( \int \mathrm{d}u \, \beta(-u) \right) 
	\prod_{r=1}^{l_n} \left( \int \mathrm{d}r' \, n(-r') \right)  \label{vertex part of the macroscopic action} \\
	& \qquad \qquad \qquad \qquad \times \mathcal{V}_{\beta\cdots\beta \, n\cdots n}(1,\dots,l_\beta,1',\dots,l_n') . \nonumber
\end{align}
Here, we introduced the inverse propagator $\Delta^{-1}$ and the $(l_\beta+l_n)$-point vertices $\mathcal{V}_{\beta\cdots\beta \, n\cdots n}$,
which can be expressed in terms of the free collective-field cumulants $G^{(0)}_{n \dotsm n B \dotsm B}$ and the one-particle potential $v$,
\begin{align}
	\Delta^{-1}(1,2) &=
	\begin{pmatrix}
	\Delta_{nn}(1,2) \;\;&\;\; \Delta_{n\beta}(1,2) \\[0.5em]
	\Delta_{\beta n}(1,2) \;\;&\;\; \Delta_{\beta\beta}(1,2) \\
	\end{pmatrix}^{-1}
	\label{inverse propagator} \\[0.5\baselineskip] &=
	\begin{pmatrix}
		v(1) \, G^{(0)}_{BB}(1,2) \, v(2) &\; \mathrm{i} \, \mathcal{I}(1,2) - v(1) \, G^{(0)}_{B n} (1,2) \\[0.5em]
		\mathrm{i} \, \mathcal{I}(1,2) - G^{(0)}_{n B} (1,2) \, v(2) & G^{(0)}_{n n} (1,2) \\
	\end{pmatrix} \!,
	\nonumber \\[0.5\baselineskip]
	\mathcal{V}_{\beta\cdots\beta \, n\cdots n}(1,\dots,l_\beta,1',\dots,l_n') &= \mathrm{i}^{l_\beta} \, (-\mathrm{i})^{l_n} \, G^{(0)}_{n \cdots n \, B \cdots B}(1,\dots,l_\beta,1',\dots,l_n') \prod_{r=1}^{l_n} v(r') ,
\end{align}
with the identity 2-point function
\begin{equation}
	\mathcal{I}(1,2)\coloneqq (2\pi)^3 \, \delta_\mathrm{D}(\vec{k}_1+\vec{k}_2) \, \delta_\mathrm{D}(t_1-t_2) .
\end{equation}}
\par
\cor{The interacting cumulants of the macroscopic fields are now obtained by taking functional derivatives of the cumulant-generating functional $W_\phi[M]\coloneqq \ln Z_\phi[M]$ with respect to the source field $M$, evaluated at $M=0$, 
\begin{align}
	G_{n\cdots n \, \beta\cdots \beta}(1,\dots,l_n,1',\dots,l_\beta') = \prod_{u=1}^{l_n} \biggl(\frac{\delta}{\mathrm{i}\delta M_n(u)} \biggr) \prod_{r=1}^{l_\beta} \biggl(\frac{\delta}{\mathrm{i}\delta M_\beta(r)} \biggr) \, W_\phi[M] \bigg|_{M=0}.
	\label{macroscopic cumulants}
\end{align}}

%%%%%%%%%%
\subsubsection*{\cor{Macroscopic Perturbation Theory}}
\cor{Similar to the treatment of the interacting part of the microscopic action in \eqref{Basics: Full Gen. Func. for Class. Fields}, the vertex part of the macroscopic action can be pulled in front of the path integral by replacing all macroscopic fields appearing in \eqref{vertex part of the macroscopic action} with functional derivatives with respect to the corresponding source fields, $\hat{S}_\mathcal{V} \coloneqq S_\mathcal{V}\bigl[\frac{\delta}{\mathrm{i} \delta M}\bigr]$. The remaining Gaussian path integral can then be performed exactly, yielding
\begin{equation}
	Z_\phi[M] = \mathrm{e}^{\mathrm{i}\hat{S}_\mathcal{V}} \, \exp\biggl\{  -\frac{1}{2} \int \mathrm{d}1 \int \mathrm{d}2 \; M^\top\!(-1) \; \Delta(1,2) \; M(-2) \biggr\} .
	\label{macroscopic generating functional with vertex operator} 
\end{equation}
Expanding the first exponential in \eqref{macroscopic generating functional with vertex operator} in orders of the vertices gives rise to the macroscopic perturbation theory of RKFT. Together with \eqref{macroscopic cumulants} this allows to write the different perturbative contributions to any macroscopic cumulant as combinations of propagators and vertices.}
\par
\cor{In this work, we specifically consider the leading-order result -- conventionally called tree-level result -- of the 2-point number density cumulant, which is precisely given by the $nn$-component of the propagator, $G^{(\mathrm{tree})}_{nn}=\Delta_{nn}$. To calculate it, a combined matrix and functional inversion of \eqref{inverse propagator}, defined via
\begin{equation}
	\int \mathrm{d}\bar{1} \; \Delta(1,\bar{1}) \; \Delta^{-1}(-\bar{1},2) = \mathcal{I}(1,2) \, \mathbbm{1}_2 
\end{equation}		
with the $2 \times 2$ identity matrix $\mathbbm{1}_2$, has to be performed. This yields
\begin{align}
	G^{(\mathrm{tree})}_{nn} = \Delta_{nn}(1,2) = G^{(0)}_{nn}(1, 2) &+ \int \mathrm{d}\bar{1} \; \tilde{\Delta}_\textsc{r}(1, -\bar{1}) \, G^{(0)}_{nn}(\bar{1}, 2) + \int \mathrm{d}\bar{2} \; G^{(0)}_{nn}(1, \bar{2}) \, \tilde{\Delta}_\textsc{a}(-\bar{2}, 2) \nonumber \\
	&+ \int \mathrm{d}\bar{1} \int \mathrm{d}\bar{2} \; \tilde{\Delta}_\textsc{r}(1, -\bar{1}) \, G^{(0)}_{nn}(\bar{1}, \bar{2}) \, \tilde{\Delta}_\textsc{a}(-\bar{2}, 2) ,
	\label{nn-component of the propagator}
\end{align}
where $\tilde{\Delta}_\textsc{r}(1,2) = \tilde{\Delta}_\textsc{a}(2,1)$ are retarded and advanced functions in time, respectively, describing the linear response of the number density at time $t_1$ to perturbations of the system at time $t_2$. They are defined as the solution of the integral equation
\begin{equation}
	\tilde{\Delta}_\textsc{r}(1, 2) = - \mathrm{i} G^{(0)}_{nB}(1, 2) \, v(2) - \int \mathrm{d}\bar{1} \; \mathrm{i} G^{(0)}_{nB}(1, \bar{1}) \, v(\bar{1}) \, \tilde{\Delta}_\textsc{r}(-\bar{1}, 2) .
	\label{integral equation for retarded propagator}
\end{equation}}
\par
\cor{For purely gravitationally interacting particles, it was shown in \cite{lilow_resummed_2019} that in the large-scale limit there exists an exact analytic solution to this equation corresponding to linear growth. But to obtain a solution that is valid on all scales and also takes into account pressure, we have to solve \eqref{integral equation for retarded propagator} numerically. To do so, we first exploit that in a statistically homogeneous situation, like cosmic structure formation, the cumulant $G^{(0)}_{nB}(1,\bar{1})$ is proportional to $\delta_\mathrm{D}(\vec{k}_1 + \vec{k}_{\bar{1}})$, rendering the integral over $\vec{k}_{\bar{1}}$ trivial. The remaining integral equation in $t_{\bar{1}}$ can then be approximated by a simple linear matrix equation by discretizing all time arguments in \eqref{integral equation for retarded propagator}. Solving this matrix equation is numerically inexpensive.}
\par
\cor{We further want to point out that there also exists a formal solution to \eqref{integral equation for retarded propagator} given by the Neumann series
\begin{equation}
\begin{split}
	\tilde{\Delta}_\textsc{r}(1, 2) = &- \mathrm{i} G^{(0)}_{nB}(1, 2) \, v(2) + \int \mathrm{d}\bar{1} \; \mathrm{i} G^{(0)}_{nB}(1, \bar{1}) \, v(\bar{1}) \, \mathrm{i} G^{(0)}_{nB}(-\bar{1}, 2) \, v(2) \\
	&- \int \mathrm{d}\bar{1} \int \mathrm{d}\bar{2} \; \mathrm{i} G^{(0)}_{nB}(1, \bar{1}) \, v(\bar{1}) \, \mathrm{i} G^{(0)}_{nB}(-\bar{1}, \bar{2}) \, v(\bar{2}) \, \mathrm{i} G^{(0)}_{nB}(-\bar{2}, 2) \, v(2) + \dotsb ,
\end{split}
\end{equation}
where the dots indicate the infinite sum of integrals over ever longer products of the function $-\mathrm{i} G^{(0)}_{nB} \, v$. While this expression does not allow to compute $\tilde{\Delta}_\textsc{r}$ explicitly, it nicely demonstrates that the leading-order RKFT result already contains contributions of arbitrarily high order in the potential $v$ and hence in the interaction operator $\hat{S}_\mathrm{I}$.}

%%%%%%%%%%%%%%%%%%%%
\subsection{Tree-Level Power Spectrum}
\label{Tree-Level Power Spectrum}
\cor{We are now in the position to computate the density contrast power spectrum \eqref{Power Spectrum} of isothermal baryons in an EdS cosmology using the RKFT formalism. For this purpose, we rewrite \eqref{Power Spectrum} as
\begin{align}
	P_\delta(1) 
	=&  \int \mathrm{d}2 \; \delta_\mathrm{D}(\eta_1-\eta_2) \, \big\langle\delta(1) \, \delta(2)\big\rangle 	\label{relation between power spectrum and nn-cumulant}\\
	=&  \frac{1}{\bar{n}^2} \int \mathrm{d}2 \; \delta_\mathrm{D}(\eta_1-\eta_2) \Big[ \big\langle n(1) \, n(2) \big\rangle - \big\langle n(1) \big\rangle \, \big\langle n(2) \big\rangle \Big] \nonumber\\
	=& \frac{1}{\bar{n}^2} \int \mathrm{d}2 \; \delta_\mathrm{D}(\eta_1-\eta_2) \, G_{nn}(1,2) , \nonumber
\end{align}
where we switched to the time coordinate $\eta=\log a$ introduced in \autoref{subsection isothermal fluid} and used the relation between 2-point cumulants and correlators in the last step. To} obtain the power spectrum we effectively have to factor \cor{$(2\pi)^3\, \delta_\mathrm{D}(\vec{k}_1+\vec{k}_2) \, \bar{n}^2$} out of the density-density cumulant $G_{nn}$ and evaluate the remainder at \cor{$\vec{k}_2=-\vec{k}_1$} as well as \cor{$\eta_2=\eta_1$}.
\par
\cor{As discussed in \autoref{Section Resummation}, at} the tree-level, i.e. at the lowest perturbative level within RKFT, $G_{nn}$ is given by the propagator component $\Delta_{nn}$ \cor{specified in \eqref{nn-component of the propagator}. The explicit expressions for the 2-point cumulants appearing in $\Delta_{nn}$ are listed in \autoref{Free Cumulants Appendix}. They are expanded only to lowest order in the initial power spectrum, as it was found in \cite{lilow_resummed_2019} that the higher-order terms only affect scales below the particles' mean free-streaming length, which we can safely neglect in the qualitative analysis presented in this paper. The one-particle potential $v$ entering the numerical evaluation of $\tilde{\Delta}_\textsc{r}$ in \eqref{integral equation for retarded propagator} is given by the sum of the effective potential contributions from pressure and gravity, $v = v_{\mathrm{p},\mathrm{eff}} + v_{\mathrm{g},\mathrm{eff}}$, defined in \eqref{effective pressure potential in EdS} and \eqref{effective gravitational potential in EdS}, respectively.}
\par
Note that this will lead us to the power spectrum based on mesoscopic particles, which is related to the power spectrum of the microscopic degrees of freedom by $P^\mathrm{mic}_\delta(k,\eta)=W^2(\vec{k})P_\delta(k,\eta)$ with $W(k)$ describing the Fourier transformed kernel function we introduced in \autoref{Section Idea of MPH}. Since we \cor{take the limit $\sigma_0\rightarrow 0$} in the following, the microscopic and the mesoscopic power spectra agree.

%%%%%%%%%%
\subsubsection*{Chosen Parameters \cor{and Initial Conditions}} 
For the initial time we choose a redshift of $z_\mathrm{i}=1100$, approximately corresponding to the time of CMB decoupling, such that $\eta_0=\ln(1+z_i)\approx 7$ corresponds to a redshift of zero.
\cor{We further use an exemplary value for the speed of sound such that $\frac{c_\mathrm{s}^2}{H_\mathrm{i}^2}=0.1 \, \mathrm{Mpc}^2 h^{-2}$ in \eqref{effective Cp}, and take the limit of a vanishing mesoscopic scale, $\sigma_0\rightarrow 0$, in all our following computations.}
\par
\cor{Accurate initial conditions for baryonic matter at the time of CMB decoupling would have to be obtained from numerical Boltzmann codes. However, we only want to investigate qualitatively how the hydrodynamic interactions affect the evolution of a pure baryon spectrum compared to the evolution of a pure dark matter spectrum. For this purpose, it is more useful to use the same initial conditions as for dark matter. In particular, this means that we are imposing the same growing mode condition $\delta^{(\mathrm{i})} = - \vec{\nabla} \cdot \vec{p}^{\;(\mathrm{i})}$ between the initial density and momentum fields as has been used in previous applications of KFT to dark matter \cite{bartelmann_microscopic_2016,bartelmann_kinetic_2017,lilow_resummed_2019}. This choice enters our calculations via the specific dependence of the free 2-point cumulants listed in \autoref{Free Cumulants Appendix} on the initial density contrast power spectrum. In our analysis, we} assume a BBKS spectrum \cite{bardeen_statistics_1986} with spectral index $n_\mathrm{s} = 1$ normalised such that $\sigma_8 = 0.8$ today.

%%%%%%%%%%
\subsubsection*{Results}
\begin{figure}
	\centering
	\includegraphics[width=0.6\textwidth]{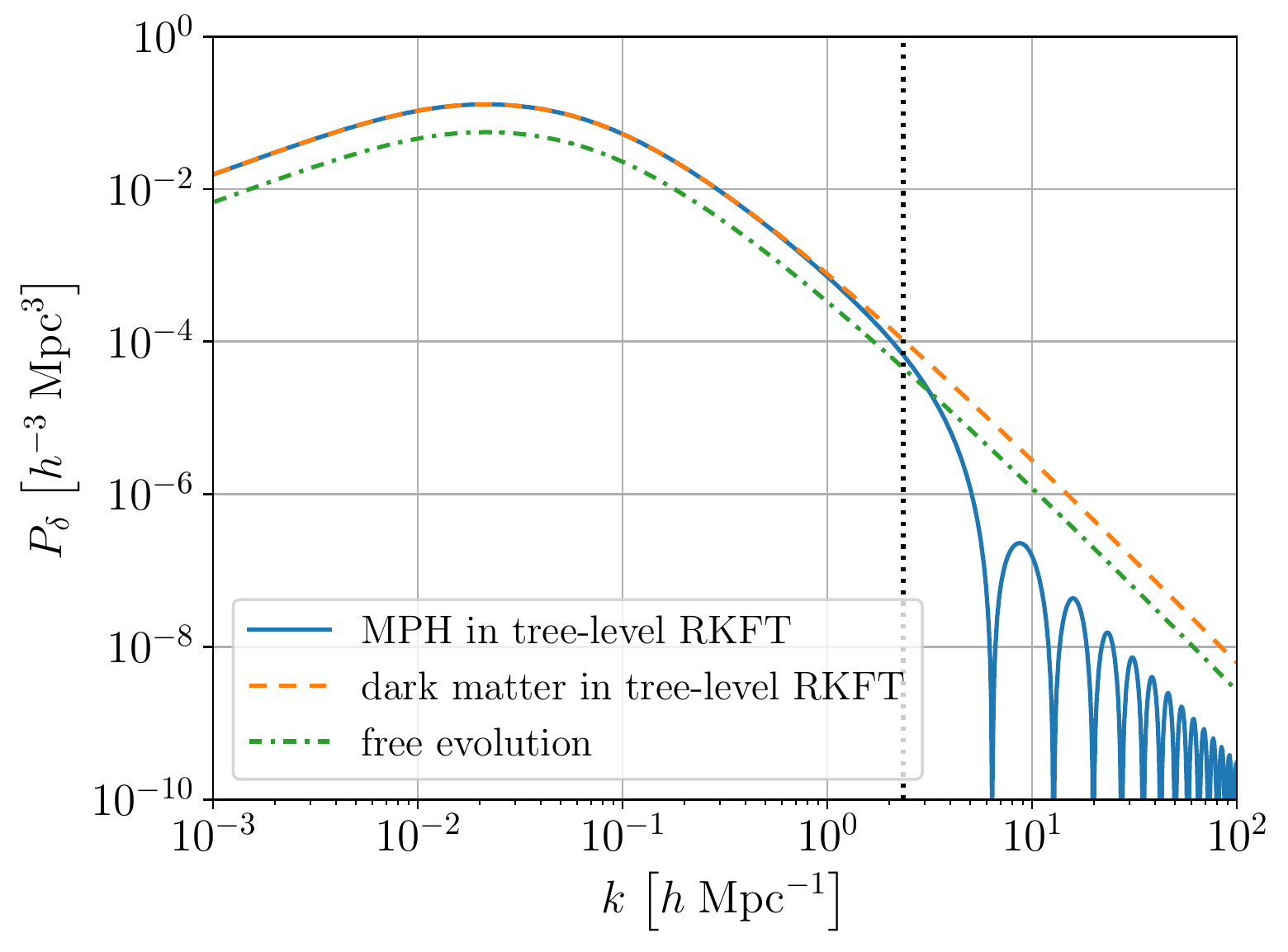}
	\caption{Comparison of the isothermal baryonic density contrast power spectrum obtained using MPH in tree-level RKFT (blue solid)\cor{, the RKFT tree-level spectrum of dark matter (orange dashed) and the spectrum obtained from freely evolving particles (green dash-dotted)} at an exemplary time $\eta=1.0$ \cor{in an EdS cosmology}. The model parameters are chosen as $\frac{c_\mathrm{s}^2}{H_\mathrm{i}^2}=0.1 \, \mathrm{Mpc}^2 h^{-2}$ and \cor{$\sigma_0 \rightarrow 0$. The wavenumber $k_{\mathrm{J},\mathrm{eff}}$ associated to the Jeans length, defined in \eqref{effective Jeans wavenumber}, is marked with a vertical dotted line. On scales larger than the Jeans length, $k \lesssim k_{\mathrm{J},\mathrm{eff}}$}, gravity dominates, leading to \cor{the same} enhanced structure growth \cor{in both baryons and dark matter when} compared to the free evolution. \cor{On smaller scales, $k \gtrsim k_{\mathrm{J},\mathrm{eff}}$, baryons} are dominated by pressure, suppressing the growth of structures \cor{below those observed in dark matter as well as freely evolving particles,} and creating oscillatory features \cor{corresponding to sound waves.}}
	\label{fig:TreeLevelMPHPowerSpectrum}
\end{figure}

The \cor{isothermal baryonic tree-level} power spectrum at an exemplary time $\eta=1.0$ is shown in \autoref{fig:TreeLevelMPHPowerSpectrum}. \cor{We compare it to the tree-level spectrum of dark matter particles, interacting only via the gravitational potential $v_{\mathrm{g},\mathrm{eff}}$, as well as the freely evolved spectrum. The latter is obtained by replacing the interacting cumulant $G_{nn}$ in \eqref{relation between power spectrum and nn-cumulant} with the free cumulant $G^{(0)}_{nn}$ given in \eqref{MPH: G_0,nn}.}
\par
From the explicit expressions for the \cor{baryonic potential contributions \eqref{effective pressure potential in EdS} and \eqref{effective gravitational potential in EdS}} we already see that for small values of $k$ gravitational effects dominate while for large $k$ pressure dominates. \cor{The transition between the gravity and pressure dominated regimes is fixed by the Jeans length. Its associated wavenumber $k_{\mathrm{J},\mathrm{eff}}$ is defined in \eqref{effective Jeans wavenumber} and marked with a vertical dotted line in \autoref{fig:TreeLevelMPHPowerSpectrum}.}
\par
\cor{In the gravity dominated regime, $k \lesssim k_{\mathrm{J},\mathrm{eff}}$, the baryonic spectrum matches the spectrum of the purely gravitationally interacting dark matter particles. Furthermore, both show a stronger growth than the freely evolved spectrum since the probability of finding matter in the neighborhood of a particle is increased by the gravitational attraction.}
\par
\cor{In the pressure dominated regime, $k \gtrsim k_{\mathrm{J},\mathrm{eff}}$, the baryonic spectrum drops below the spectra of both dark matter and freely evolving particles,} due to the repulsive nature of pressure. In addition, we observe an oscillating behavior which describes the phenomenon of sound waves. The appearance of sharp edges corresponds to the fact that the density fluctuations may have negative values for some $k$ which cancel out in the calculation of the power spectrum by taking the square. By choosing a higher resolution one indeed finds that these edges approach \cor{zero.}
\par
If one would choose some finite value for the mesoscopic scale, the model will break down for wavenumbers $k \gtrsim \pi/\sigma_0$. In this regime the resulting power spectrum approaches the free spectrum. This behavior can easily be explained by the construction of the model: Consider two mesoscopic particles labeled by $i$ and $j$ in a very small distance $\vec{r}\rightarrow0$ to each other such that they strongly overlap in their microscopic structure. The total gravitational force acting on the particle $i$ vanishes since the microscopic particles of $j$ symmetrically pull it in each direction. Similarly, the force caused by a pressure gradient vanishes due to the symmetry of the kernel function $W_j$.

%%%%%%%%%%%%%%%%%%%%%%%%%%%%%%%%%%%%%%%%
\section{Eulerian Perturbation Theory}
\label{Section EPT}
In this section, we show how baryonic matter can be treated in the framework of Eulerian perturbation theory, which is currently the standard approach for analytical description of cosmic structure formation, and compare the outcome with our findings from the resummed MPH model.
\par
Typically, the Eulerian approach is used to model only dark matter. For this purpose, it is assumed that dark matter can approximately be described by a pressureless perfect fluid\footnote{This is only a rough approximation since dark matter will form multiple streams, which is not captured in the perfect fluid description. In contrast, in the framework of KFT the trajectories of particles can cross such that this effect is included by construction.} such that the hydrodynamical equations hold. The construction is based on the assumption that mass and momentum of the matter content is homogeneously distributed on large scales such that the hydrodynamical equations can be expanded in small perturbations around the mean values. For consideration of baryonic degrees of freedom, pressure has to be taken into account additionally. In the following, we focus on the linearized equations only, to compare it with our results above.

%%%%%%%%%%%%%%%%%%%%
\subsection{Setting up the Equations of Motion}
\label{Section: EPH EoM's}
As before, we consider an isothermal fluid in an expanding universe. Following the standard literature, we use comoving coordinates $\vec{q}$ and the conformal time $\tau$ which is related to the cosmic time by $\mathrm{d}t = a(\tau)\mathrm{d}\tau$. For the velocity field $\vec{u}$ we write
\begin{equation}
	\vec{u}(\vec{q},\tau) \coloneqq \mathcal{H}(\tau)\vec{q} + \vec{u}_\mathrm{p}(\vec{q},\tau),
\end{equation}
where the first term describes the Hubble flow with $\mathcal{H}\coloneqq\mathrm{d}\ln a/\mathrm{d}\tau$ and the second is known as the peculiar velocity of the fluid, $\vec{u}_\mathrm{p}=\mathrm{d}\vec{q}/\mathrm{d}\tau$. Furthermore, $\delta(\vec{q},\tau)$ describes the density contrast we already introduced in the last section. Using all these expressions one can show that the continuity, Euler and Poisson equations take the form
\begin{align}
	& \partial_\tau \delta + \nabla_q \big( (\delta+1)\vec{u}_\mathrm{p}\big) = 0	,	\\
	& \partial_\tau \vec{u}_\mathrm{p} + \frac{1}{\rho}\nabla_q P + \mathcal{H}\vec{u}_\mathrm{p} + \big(\vec{u}_\mathrm{p}\cdot\nabla_q \big) \vec{u}_\mathrm{p} = -\nabla_q \Phi_\mathrm{g}	,	\\
	&\nabla_q^2 \Phi_\mathrm{g} = \frac{3}{2}\mathcal{H}^2 \Omega_\mathrm{m} \delta
\end{align}
with
\begin{equation}
	\Phi_\mathrm{g} \coloneqq \frac{1}{2} \partial_\tau \vec{q}^2 + \phi_\mathrm{g}.
\end{equation}
For a detailed derivation see for example \cite{bernardeau_large-scale_2002}. Note the appearance of an additional term corresponding to pressure effects. Inserting the isothermal equation of state \eqref{isothermal_equation_of_state} and expanding the factor $1/\rho$ one finds
\cor{\begin{equation}
	\frac{1}{\rho}\nabla_q P = c_\mathrm{s}^2 \sum_{n=0}^{\infty} (-\delta)^n \nabla_q \delta.
\end{equation}}
To avoid vectorial degrees of freedom, we take the divergence of the modified Euler equation and introduce the velocity divergence $\Theta\coloneqq \nabla_q\cdot \vec{u}_\mathrm{p}$. Assuming an irrotational fluid, its evolution is fully described in terms of only $\delta$ and $\Theta$.
\par
Going to Fourier space, the equations of motion read
\cor{\begin{align}
	\partial_\tau \tilde{\delta}(\vec{k},\tau) &+ \tilde{\Theta}(\vec{k},\tau) = - \!\! \int \!\! \frac{\mathrm{d}^3k_1}{(2\pi)^3} \!\! \int \!\! \frac{\mathrm{d}^3k_2}{(2\pi)^3} \, (2\pi)^3  \delta_\mathrm{D}(\vec{k}-\vec{k}_1-\vec{k}_2) \alpha(\vec{k}_1,\vec{k}_2)\tilde{\Theta}(\vec{k}_1,\tau) \tilde{\delta}(\vec{k}_2,\tau) ,	\label{EPH: EOM1 Fourier}	\\
	\partial_\tau \tilde{\Theta}(\vec{k},\tau) &- c_\mathrm{s}^2 k^2 \tilde{\delta}(\vec{k},\tau) +  \mathcal{H}(\tau)\tilde{\Theta}(\vec{k},\tau) + \frac{3}{2}\Omega_\mathrm{m}(\tau)\mathcal{H}(\tau)^2 \tilde{\delta}(\vec{k},\tau)   \nonumber	\\
	= &- \!\! \int \!\! \frac{\mathrm{d}^3k_1}{(2\pi)^3} \!\! \int \!\! \frac{\mathrm{d}^3k_2}{(2\pi)^3} \, (2\pi)^3  \delta_\mathrm{D}(\vec{k}-\vec{k}_1-\vec{k}_2) \beta(\vec{k}_1,\vec{k}_2) \tilde{\Theta}(\vec{k}_1,\tau) \tilde{\Theta}(\vec{k}_2,\tau)   \label{EPH: EOM2 Fourier}   \\
	&- \sum_{n=2}^\infty \; \prod_{r=1}^n \, \biggl(\int \!\! \frac{\mathrm{d}^3k_r}{(2\pi)^3} \, \tilde{\delta}(\vec{k}_r,\tau)\biggr) \, (2\pi)^3 \delta_\mathrm{D}\bigg(\vec{k} - \sum_{r=1}^n \vec{k}_r\bigg) \, \mu_n(\vec{k}_1, \dotsc, \vec{k}_n) , \nonumber
\end{align}}
with
\begin{align}
	\alpha(\vec{k}_1,\vec{k}_2) &\coloneqq \frac{(\vec{k}_1+\vec{k}_2)\cdot \vec{k}_1}{k_1^2},		\\
	\beta(\vec{k}_1,\vec{k}_2) &\coloneqq \frac{(\vec{k}_1 + \vec{k}_2)^2 (\vec{k}_1\cdot\vec{k}_2)}{2k_1^2k_2^2},		\\
	\cor{\mu_n(\vec{k}_1, \dotsc, \vec{k}_n)} &\cor{\coloneqq (-1)^n \, \frac{c_\mathrm{s}^2}{n} \, \bigg(\sum_{r=1}^n \vec{k}_r\bigg)^2.}
\end{align}
The two equations above can be written in a more compact form by defining a doublet field $\varphi(\vec{k},\eta) \coloneqq (\tilde{\delta}(\vec{k},\eta), - \tilde{\Theta}(\vec{k},\eta)/\mathcal{H(\eta)})$ where $\eta = \ln a$ is the dimensionless time variable we already defined earlier. Using the \cor{conventions that repeated Fourier arguments are integrated over and repeated indices $a,b,b_r \in \{1,2\}$ are summed over}, the equation of motion becomes
\cor{\begin{equation}
	\Big(\partial_\eta \delta_{ab} + M_{ab}(\eta) \Big)\varphi_b(\vec{k},\eta) = \sum_{n=2}^\infty \, \gamma_{a b_1 \dotsb b_n} \, (\vec{k},\vec{k}_1, \dots, \vec{k}_n;\eta) \, \prod_{r=1}^n \,  \varphi_{b_r}(\vec{k}_r,\eta) ,
	\label{EPH: EOM Eff Fluid}
\end{equation}}
where
\begin{equation}
	M(\eta) \coloneqq
	\begin{pmatrix}
		0 & -1 \\
		-\frac{3}{2}\Omega_\mathrm{m}(\eta) + \frac{c_\mathrm{s}^2k^2}{\mathcal{H}(\eta)^2} \;\;&\;\; \frac{1}{2}
	\end{pmatrix}
\end{equation}
and $\gamma$ is the symmetrized vertex \cor{tensor} with the only non-vanishing elements given by
\cor{\begin{align}
	\gamma_{121} (\vec{k}, \vec{k}_1,\vec{k}_2;\eta) &= \delta_\mathrm{D}(\vec{k}-\vec{k}_1-\vec{k}_2) \, \frac{\alpha(\vec{k}_1,\vec{k}_2)}{2} ,  \\
	\gamma_{112} (\vec{k}, \vec{k}_1,\vec{k}_2;\eta) &= \delta_\mathrm{D}(\vec{k}-\vec{k}_1-\vec{k}_2) \, \frac{\alpha(\vec{k}_2,\vec{k}_1)}{2} ,   \\
	\gamma_{222} (\vec{k}, \vec{k}_1,\vec{k}_2;\eta) &= \delta_\mathrm{D}(\vec{k}-\vec{k}_1-\vec{k}_2) \, \beta(\vec{k}_1,\vec{k}_2) ,   \\
	\gamma_{2 1 \dotsb 1} (\vec{k}, \vec{k}_1, \dotsc, \vec{k}_n; \eta) &= \delta_\mathrm{D}\bigg(\vec{k} - \sum_{r=1}^n \vec{k}_r\bigg) \, \frac{\mu_n(\vec{k}_1, \dotsc, \vec{k}_n)}{\mathcal{H}(\eta)^2} .
\end{align}}

%%%%%%%%%%%%%%%%%%%%
\subsection{Solution of the Linearized Equations}
To calculate the linear power spectrum we have to determine the Green's function of the linear part of equation \eqref{EPH: EOM Eff Fluid}. The calculation can be done analogously to the pressureless case which can be found in Ref. \cite{crocce_renormalized_2006}. For the general case with a time-dependent matrix $M(\eta)$ there is no analytical solution known. Thus, to keep our analysis simple, we assume an Einstein-de Sitter universe (i.e. $\Omega_\mathrm{m}=1$ and $H=H_\mathrm{i}a^{-3/2}$) and approximate the time-dependent term \cor{$\frac{c_\mathrm{s}^2}{\mathcal{H}(\eta)^2}$} by its mean value
\cor{\begin{equation}
	\ell^2 \coloneqq \frac{1}{\eta_\mathrm{f}-\eta_\mathrm{i}} \int_{\eta_\mathrm{i}}^{\eta_\mathrm{f}} \mathrm{d}\eta \frac{c_\mathrm{s}^2}{\mathcal{H}(\eta)^2} = \frac{1}{\eta_\mathrm{f}-\eta_\mathrm{i}} \frac{c_\mathrm{s}^2}{H_\mathrm{i}^2} \bigg( \mathrm{e}^{\eta_\mathrm{f}} - \mathrm{e}^{\eta_\mathrm{i}} \bigg),
	\label{averaging_the_time_dependence_in_the_EPT_pressure_term}
\end{equation}}
where $[\eta_\mathrm{i},\eta_\mathrm{f}]$  denotes the time interval we are interested in. This approximation obviously gets better as we choose a very small time interval. Therefore, for high accuracy it is advisable to split the interval in smaller subintervals and perform calculations step by step.
\par
Since all time dependencies of the equation are \cor{now} inhabited in the fields $\varphi_a$ only, we are able to perform a Laplace transform \cor{of the linear part of} \eqref{EPH: EOM Eff Fluid}. Defining $\varphi(\vec{k},\omega)\coloneqq\mathcal{L} \big\{\varphi(\vec{k},\eta) \big\} (\omega)$, this leads us to
\cor{\begin{equation}
	\sigma_{ab}^{-1}(\omega)\varphi^{\mathrm{L}}_b(\vec{k},\omega) = \varphi_a^{(\mathrm{i})}(\vec{k}) ,		\label{EPH: EOM Eff Fluid Laplace}
\end{equation}}
where \cor{$\varphi_a^{(\mathrm{i})}(\vec{k}) \coloneqq \varphi_a(\vec{k},\eta_\mathrm{i})$} and $\sigma^{-1}_{ab}(\omega)\coloneqq \omega\delta_{ab} + M_{ab}$ with the inverse matrix 
\begin{equation}
	\sigma_{ab}(\omega) = \frac{1}{(\omega-\omega_1)(\omega - \omega_2)}
	\begin{pmatrix}
		\omega + \frac{1}{2} & 1 \\
		\frac{3}{2} - k^2\ell^2 & \omega\\
	\end{pmatrix}.
\end{equation}
The poles are given by $\omega_{1/2}(k) \coloneqq \pm \alpha(k) -\frac{1}{4}$ and \cor{$\alpha(k) \coloneqq \sqrt{\frac{25}{16} - k^2\ell^2}$}.
\par
Multiplying equation \eqref{EPH:  EOM Eff Fluid Laplace} with $\sigma_{ab}$ and inverting the Laplace transform, we obtain 
\cor{\begin{equation}
	\varphi^{\mathrm{L}}_a(\vec{k},\eta) = g^\mathrm{R}_{ab}(\vec{k};\eta,\eta_\mathrm{i}) \varphi_b^{(\mathrm{i})}(\vec{k})
\end{equation}}
with the Green's function
\cor{\begin{align}
	g^\mathrm{R}_{ab}(\vec{k};\eta,\eta_\mathrm{i}) &= \oint_{c-\mathrm{i}\infty}^{c+\mathrm{i}\infty} \frac{\mathrm{d}\omega}{2\pi \mathrm{i}}\ \sigma_{ab}(\omega) \mathrm{e}^{\omega(\eta-\eta_\mathrm{i})}  \label{EPH: GF}\\
	&= \frac{\mathrm{e}^{-\Delta\eta/4}}{\alpha}
	\begin{pmatrix}
		\alpha \cosh(\alpha \Delta\eta) + \frac{1}{4} \sinh(\alpha\Delta\eta) & \sinh(\alpha \Delta\eta) \\
		(\frac{3}{2} - k^2\ell^2) \sinh(\alpha \Delta\eta) & \alpha \cosh(\alpha \Delta\eta ) - \frac{1}{4} \sinh(\alpha \Delta\eta)\\
	\end{pmatrix} 
	\theta(\Delta\eta).  \nonumber
\end{align}}
For better readability, the $k$-dependence of the frequency $\alpha(k)$ is suppressed and we used the abbreviation $\Delta\eta \coloneqq \eta-\eta_\mathrm{i}$. From this expression we can infer some insight about  the underlying physics. For large distances the frequency $\alpha(k)$ becomes real and hence we obtain some hyperbolic behavior, while for $k^2>\frac{25}{16}\ell^{-2}$ the hydrodynamic fields oscillate. This threshold corresponds to the Jeans length modified by a model specific shift of $+\frac{1}{16}\ell^{-2}$.
\par
Considering the limit of a pressureless fluid, corresponding to $c_\mathrm{s}\rightarrow0$, we find the propagator to be
\cor{\begin{equation}
	\tilde{g}^\mathrm{R}_{ab} (\eta,\eta_\mathrm{i}) = \frac{\mathrm{e}^{\Delta\eta}}{5} \begin{pmatrix}
		3 & 2 \\
		3 & 2 \\
	\end{pmatrix}
	- \frac{\mathrm{e}^{-3/2\Delta\eta}}{5}
	\begin{pmatrix}
		-2 & 2 \\
		3 & -3 \\
	\end{pmatrix}  \label{modes}
\end{equation}}
which is identical to the result of \cite{crocce_renormalized_2006}. From this expression one can easily read of the growing  and decaying modes to be proportional to
\begin{align}
	u_a= 
	\begin{pmatrix}
	1 \\ 1
	\end{pmatrix}
	\qquad
	\text{and}
	\qquad
	v_a=
	\begin{pmatrix}
	1 \\ -3/2
	\end{pmatrix} \label{EPH: Mode Decomposition},
\end{align}
respectively.
\par
In a model including pressure effects there exist no such universal vectors describing global growing and decaying modes due to the oscillating behavior on small scales. Nevertheless, on large scales, where such effects can be \cor{neglected}, a decomposition in growing and decaying modes is still applicable. \cor{Beyond that, we deliberately used large-scale growing mode initial conditions in our calculations of the MPH power spectrum in \autoref{Tree-Level Power Spectrum} since we only wanted to investigate the impact of the different dynamics in baryonic and dark matter, and not the influence of their different initial conditions. For a consistent comparison between the MPH and EPT approaches, we thus also have to use the large-scale growing mode solution here.}
\par
With the Green's function we can calculate the linear power spectrum $P^\mathrm{L}$ from the initial power spectrum \cor{$P^\mathrm{(i)}$} by
\begin{align}
	P^\mathrm{L}_{ab}(k,\eta) = g_{ac}^\mathrm{R}(k;\eta,0) g_{bd}^\mathrm{R}(k;\eta,0) P_{cd}^{(\mathrm{i})}(k) \label{EPH: P^L relation} , 
\end{align}
with $P^{\mathrm{X}}_{ab}\coloneqq u_a u_b P_\delta^{\mathrm{X}}$.

%%%%%%%%%%%%%%%%%%%%
\subsection{Comparison with Resummed MPH}
\begin{figure}
	\centering
	\includegraphics[width=0.6\textwidth]{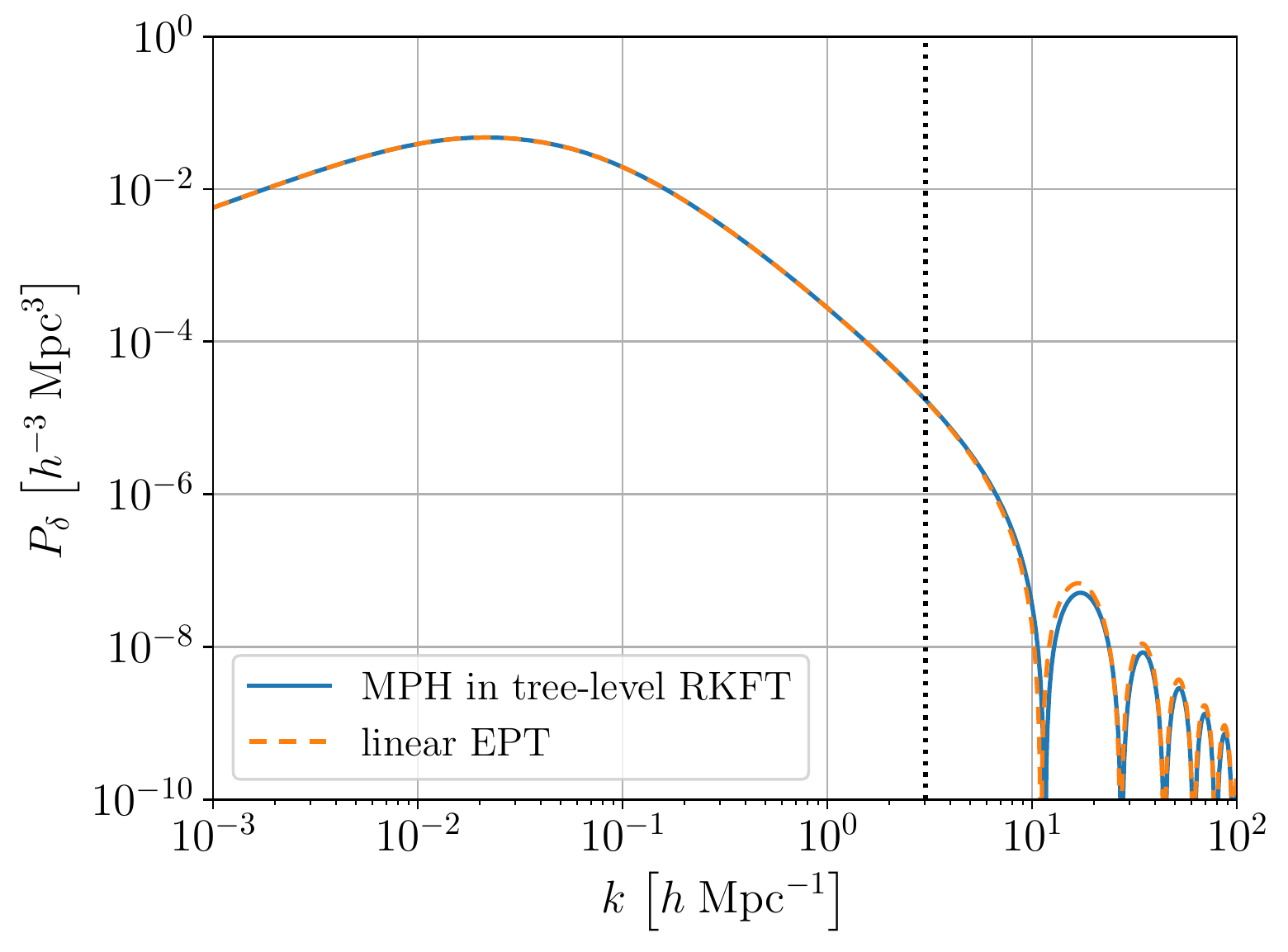}
	\caption{Comparison of the isothermal baryonic density contrast power spectrum obtained using MPH in tree-level RKFT (blue solid) and the respective spectrum obtained from linear EPT (orange dashed) at an exemplary time $\eta=0.5$. The model parameters are chosen as $\frac{c_\mathrm{s}^2}{H_\mathrm{i}^2}=0.1 \, \mathrm{Mpc}^2 h^{-2}$ and \cor{$\sigma_0 \rightarrow 0$. The wavenumber $k_{\mathrm{J},\mathrm{eff}}$ associated to the Jeans length, defined in \eqref{effective Jeans wavenumber}, is marked with a vertical dotted line.} On large, gravity-dominated scales\cor{, $k \lesssim k_{\mathrm{J},\mathrm{eff}}$,} both spectra agree perfectly, while slight deviations can be seen on small, pressure-dominated scales \cor{below the Jeans length, $k \gtrsim k_{\mathrm{J},\mathrm{eff}}$}. The latter is a consequence of the approximation made in the EPT calculation by averaging the time-dependence of the pressure term in \eqref{averaging_the_time_dependence_in_the_EPT_pressure_term}.}
	\label{fig:CompareMPHAndEPTPowerSpectra}
\end{figure}

With this EPT approach for baryonic matter, we are now in the position to quantitatively verify our findings from the resummed MPH model. As before, we consider an EdS universe  with an initial BBKS spectrum at redshift $z_\mathrm{i}=1100$ and set $\frac{c_\mathrm{s}^2}{H_\mathrm{i}^2}=0.1 \, \mathrm{Mpc}^2 h^{-2}$. The resulting power spectra at an exemplary time $\eta=0.5$ are plotted in \autoref{fig:CompareMPHAndEPTPowerSpectra} and show an excellent agreement, in particular on large scales.
\par
On small scales there are slight deviations which are not unexpected. In the EPT approach we averaged the term $\frac{c_\mathrm{s}^2}{\mathcal{H}(\eta)^2}$ to be able to perform a Laplace transform. Since this term appears in context with pressure effects it is reasonable that the approximation influences the small scale behavior and hence explains the discrepancy to the resummed MPH model. Considering later times, the effects due to this time averaging will grow and thus the deviations do so as well. In the resummed MPH model we did not make such an approximation, therefore it is reasonable to assume that the MPH results are more accurate. To improve the model of EPT one could implement the calculation of the Green's function numerically, which would render the averaging unnecessary. However, this is not the aim of this paper.
\par
Furthermore, the agreement between the results from MPH in tree-level RKFT, which contains terms of arbitrarily high order in the mesoscopic interaction operator, and linear EPT implies that an expansion to any finite order in the interaction operator would not suffice to fully reproduce the solution of even the linearized hydrodynamical equations. Thus, a consistent description of fluid dynamics within KFT indeed requires to use the RKFT framework, as already suspected in \cite{viermann_model_2018}.

%%%%%%%%%%%%%%%%%%%%%%%%%%%%%%%%%%%%%%%%
\section{Conclusion and Outlook}
\label{Section Conclusion}
In this paper we used the model of Mesoscopic Particle Hydrodynamics (MPH), introduced in \cite{viermann_model_2018}, to tackle the difficulties of including baryonic degrees of freedom in the analysis of cosmic structure formation. MPH considers effective particles of finite extent whose dynamics follow the hydrodynamical equations and describes them within the framework of Kinetic Field Theory (KFT). Here, we presented an alternative derivation of the mesoscopic equations of motion, which verified the findings of \cite{viermann_model_2018} and also allowed for an easier generalization of the model to a cosmological setting. 
\par
Subsequently, we focused on the simplified case of an isothermal fluid, which is fully characterized by the free motion of the mesoscopic particles plus an interaction potential consisting of contributions from pressure and gravity. To make it applicable for the issue of cosmic structure formation we translated the potential to an expanding space-time. 
\par
We performed a resummation of the mesoscopic particle interactions using the RKFT formalism of Lilow et al. \cite{lilow_resummed_2019} and calculated the density power spectrum for an EdS universe to lowest order in RKFT perturbation theory. Already at this low level we obtained qualitatively good results which agree with the expectations: \cor{On} large scales gravitational effects become dominant, leading to \cor{precisely the same growth of structures as found in dark matter. Small scales below the Jeans length, on the other hand,} are dominated by pressure effects\cor{, causing a suppression of structure growth compared to dark matter as well as} an oscillatory behavior of the power spectrum corresponding to sound waves. For a quantitative validation of resummed MPH, we compared the model with Eulerian perturbation theory (EPT) and found good agreement of the resulting power spectra.
\par
With the model of resummed MPH at hand, we are now in a position to handle both dark and baryonic matter in the framework of KFT. While a path integral description of baryonic matter is also possible in the fluid picture of EPT, coupling it to the dark matter particles of KFT would be very cumbersome if even possible. Thus, for a treatment of the full cosmic matter content within one analytic framework, the particle-based model of resummed MPH is the preferred approach. A subsequent paper will explicitly discuss the coupling of two particle species in KFT and show how to resum the combined theory of dark and baryonic matter accordingly. 
\par
Lastly, we want to stress that MPH is not restricted to the special case of an isothermal fluid. By taking the enthalpy of a mesoscopic particle fully into account and including the corresponding equation of motion in the KFT formalism, one can also describe general ideal or even viscous fluid dynamics, as demonstrated in \cite{viermann_model_2018}. The resummation of more realistic baryonic fluid dynamics will be the subject of upcoming work.

\acknowledgments{
    We are grateful for many helpful comments and discussions to Celia Viermann and Elena Kozlikin. This work was supported in part by the German Excellence Initiative, the Heidelberg Graduate School of Physics, the Collaborative Research Centre TRR 33 ‘The Dark Universe’ of the German Research Foundation, the International Max Planck Research Schools (IMPRS) and by a Technion fellowship.
}

\appendix

%%%%%%%%%%%%%%%%%%%%%%%%%%%%%%%%%%%%%%%%
\section{Derivation of the Equations of Motion for Mesoscopic Particles}
\label{EOM's Appendix}
In this section we give a more detailed derivation of the equations of motion according to \autoref{EOM's Main}.

%%%%%%%%%%
\subsubsection*{Euler Equation}
We consider the $j$-th particle in its rest frame. To obtain the contributions on a single particle, we multiply the hydrodynamic equations by a projection function $\Pi_j(\vec{r})$ and integrate over $\mathbb{R}^3$. We exemplified this procedure on the basis of the first term. For the second term we use equation \eqref{enthalpy} and insert the expression for the density \eqref{density}, then
\begin{align}
	\int \mathrm{d}^3r\ \Pi_j(\vec{r}) \nabla_{r} P(\vec{r})  \label{EOMforce}
	& =\int \mathrm{d}^3r\ \Pi_j(\vec{r}) \nabla_{r} \bigg(\sum_{i} \frac{\gamma-1}{\gamma} H_{i} W_i(\vec{r})\bigg)	\\
	& = \sum_{i}   \frac{\gamma-1}{\gamma} H_{i} \int \mathrm{d}^3r\ \Pi(\vec{r}-\vec{r}_j) \nabla_{r} W(\vec{r}-\vec{r}_i) 	\nonumber\\
	& = - \sum_{i}   \frac{
		\gamma-1}{\gamma} H_{i} \nabla_{r_i}\int \mathrm{d}^3r\ \Pi(\vec{r}-\vec{r}_j)  W(\vec{r}-\vec{r}_i) 	\nonumber\\
	& = \nabla_{r_j} \sum_{i}   \frac{\gamma-1}{\gamma}H_{i} \int \mathrm{d}^3r\ \Pi\big(\vec{r}\big)  W\big(\vec{r}-(\vec{r}_i-\vec{r}_j)\big) 	\nonumber\\
	& \eqqcolon  M \nabla_{r_j} V_\mathrm{p}(r_j)	\nonumber,
\end{align}
where we used in the fourth line that terms in which $i=j$ correspond to self-interactions, which vanish due to symmetry, and introduced the mesoscopic pressure potential $V_\mathrm{p}$ in the last line.
\par
To include gravitational effects between the underlying particles, we consider the conventional Newtonian potential
\begin{equation}
	\phi_\mathrm{g}(\vec{r})=\int \mathrm{d}^3 r' \frac{G}{|\vec{r}-\vec{r}'|} \rho(\vec{r}').
\end{equation}
Inserting this into the Euler equation, the contributions of the last term become
\begin{align}
	\int \mathrm{d}^3r\ \Pi_j(\vec{r})\rho(\vec{r}) \nabla_r \phi(\vec{r}) &= 	\int \mathrm{d}^3r\ \Pi_j(\vec{r})\sum_i M W_i(\vec{r}) \nabla_r \phi_\mathrm{g}(\vec{r})  \\
	&=  \int \mathrm{d}^3r\ \Pi_j(\vec{r}) M W_j(\vec{r}) \nabla_r \phi_\mathrm{g}(\vec{r})  \nonumber\\
	&= \int \mathrm{d}^3r\ \Pi(\vec{r}-\vec{r}_j) W(\vec{r}-\vec{r}_j) \nabla_r \bigg(\int \mathrm{d}^3 r' \frac{GM^2}{|\vec{r}-\vec{r}'|} \sum_i W(\vec{r}'- \vec{r}_i) \bigg)   \nonumber\\
	&= \int \mathrm{d}^3r''\ \Pi(\vec{r}'') W(\vec{r}'') \nabla_{r''} \bigg(\int \mathrm{d}^3 r' \frac{GM^2}{|\vec{r}''+\vec{r}_j-\vec{r}'|} \sum_i W(\vec{r}'- \vec{r}_i) \bigg)   \nonumber\\
	&= \nabla_{r_j} \sum_i \int \mathrm{d}^3 r' \mathrm{d}^3r''\ \frac{GM^2}{|\vec{r}''+\vec{r}_j-\vec{r}'|} \Pi(\vec{r}'') W(\vec{r}'')    W(\vec{r}'- \vec{r}_i)    \nonumber\\
	&\eqqcolon M \nabla_{r_j} V_\mathrm{g}(\vec{r}_j), \nonumber
\end{align}
where we used in the second line that all terms where $i\neq j$ vanish identically, which can be seen as follows,
\begin{align}
	&\sum_{i\neq j} \int \mathrm{d}^3r\  \Pi(\vec{r}-\vec{r}_j) W(\vec{r}-\vec{r}_i) \nabla_r \phi_\mathrm{g} (\vec{r}) \\ =&- \sum_{i\neq j} \int \mathrm{d}^3r\  \phi_\mathrm{g} (\vec{r}) \nabla_r \bigg[ \Pi(\vec{r}-\vec{r}_j) W(\vec{r}-\vec{r}_i)\bigg] \nonumber\\
	=&\sum_{i\neq j} \bigg( \nabla_{r_j} +\nabla_{r_i} \bigg) \int \mathrm{d}^3r\  \phi_\mathrm{g} (\vec{r}) \bigg[ \Pi(\vec{r}-\vec{r}_j) W(\vec{r}-\vec{r}_i)\bigg] \nonumber\\
	=&\sum_{i\neq j} \bigg( \nabla_{r_j} +\nabla_{r_i} \bigg) \int \mathrm{d}^3r'\  \phi_\mathrm{g} (\vec{r}') \bigg[ \Pi(\vec{r}') W(\vec{r}'+\vec{r}_j-\vec{r}_i)\bigg] \nonumber\\
	=&\sum_{i\neq j} \bigg( \nabla_{r_j} -\nabla_{r_j} \bigg) \int \mathrm{d}^3r'\  \phi_\mathrm{g} (\vec{r}') \bigg[ \Pi(\vec{r}') W(\vec{r}'+\vec{r}_j-\vec{r}_i)\bigg] \nonumber\\
	=& 0 .\nonumber
\end{align}

%%%%%%%%%%
\subsubsection*{Energy Equation}
To derive the equation of motion corresponding to the enthalpy, we have to massage the energy equation \eqref{MPH: EnergyEq} a bit,
\begin{align}
	&\frac{\mathrm{d}}{\mathrm{d}t} \varepsilon + (\varepsilon+P) \nabla_r \cdot  \vec{u} =0   \nonumber\\
	\Leftrightarrow\ & n\frac{\mathrm{d}}{\mathrm{d}t}E + E\frac{\mathrm{d}}{\mathrm{d}t}n +  \nabla_r\cdot(h\vec{u}) -\vec{u} \cdot \nabla_r h = 0  .\nonumber
\end{align}
Applying the same steps as for the Euler equation, the first term becomes
\begin{align}
	\int \mathrm{d}^3r\  \Pi_j(\vec{r}) n(\vec{r}) \frac{\mathrm{d}}{\mathrm{d}t}E(\vec{r}) \approx \frac{1}{\gamma}\frac{\mathrm{d}}{\mathrm{d}t} H_j.   
\end{align}
The second term corresponds to the dynamics of the kernels $W_i$ and is given by
\begin{align}
	\int \mathrm{d}^3r\ \Pi_j(\vec{r}) E \frac{\mathrm{d}}{\mathrm{d}t}n(\vec{r}) = &
	\int \mathrm{d}^3r\  \Pi_j(\vec{r})  \sum_i E_i \vec{u}_i \cdot \nabla_{r_i} W_i(\vec{r}) \\ =& -\sum_i \frac{1}{\gamma} H_i\vec{u}_i \cdot \nabla_{r_j} \int \mathrm{d}^3r\  \Pi\big(\vec{r}\big)W\big(\vec{r}-(\vec{r}_i-\vec{r}_j)\big).  \nonumber
\end{align}
Since we consider the rest frame of the $j$-th particle, the last term is approximately zero while the third term contributes
\begin{align}
	\int \mathrm{d}^3r\ \Pi_j(\vec{r}) \nabla_r \cdot \big(h\vec{u}(\vec{r})\big) &= \int \mathrm{d}^3r\ \Pi_j(\vec{r}) \nabla_r \cdot \sum_i H_i \vec{u}_i W_i(\vec{r})    \\
	&= \sum_i H_i \vec{u}_i \cdot \nabla_{r_j }\int \mathrm{d}^3r\ \Pi\big(\vec{r}\big)   W\big(\vec{r}-(\vec{r}_i-\vec{r}_j)\big).   \nonumber
\end{align}
Putting everything together and transforming back to a general coordinate system, we end up \cor{with \eqref{mes energy equation}.}

%%%%%%%%%%%%%%%%%%%%%%%%%%%%%%%%%%%%%%%%
\section{Explicit Calculation of the Fourier transformed Gravitational Potential}
\label{Potentials Appendix}
While the pressure potential simply corresponds to an evaluation of a convolution, the calculation of the gravitational potential is a bit more cumbersome. To save some breath we will directly determine the Fourier transform of $v_\mathrm{g}$. But let us start by first evaluating the $\vec{q}_2$-integration in \eqref{V_g},
\begin{align}
	v_\mathrm{g}(\vec{R}) &= GM \int \mathrm{d}^3 q_1 \mathrm{d}^3q_2\ \frac{1}{|\vec{q}_2-\vec{q}_1|} \Pi\big(\vec{q}_2\big) W(\vec{q}_2)    W\big(\vec{q}_1- \vec{R}\big) \\
	&= GM \int \mathrm{d}^3 q_1\  \frac{1}{|\vec{q}_1|}   
	\int  \mathrm{d}^3q_2\  \Pi\big(\vec{q}_2\big) W(\vec{q}_2)    W\big(\vec{q}_1 + \vec{q}_2 - \vec{R}\big)	\nonumber \\
	&= GM A^2B \bigg(\frac{2\pi \sigma^2}{3} \bigg)^{3/2} \int \mathrm{d}^3 q_1\  \frac{1}{|\vec{q}_1|}  \exp\bigg\{-\frac{(\vec{q}_1-\vec{R})^2}{3\sigma^2}\bigg\}	.	\nonumber
\end{align}
Before we execute the last spatial integration we perform a Fourier transform,
\begin{align}
	v_\mathrm{g}(\vec{k}) &= \int \mathrm{d}^3R\  \mathrm{e}^{\mathrm{i}\vec{k}\cdot\vec{R}} v_\mathrm{g}(\vec{R})   \\
	&=  GM A^2B \bigg(\frac{2\pi \sigma^2}{3} \bigg)^{3/2} \int \mathrm{d}^3 q_1\  \frac{1}{|\vec{q}_1|} \ \int \mathrm{d}^3R\  \mathrm{e}^{\mathrm{i}\vec{k}\cdot\vec{R}} \exp\bigg\{-\frac{(\vec{q}_1-\vec{R})^2}{3\sigma^2}\bigg\}		\nonumber\\
	&=  GM A^2B \big(2\pi^2\sigma^4 \big)^{3/2} \int \mathrm{d}^3 q_1\  \frac{1}{|\vec{q}_1|} \exp\bigg\{-\frac{3}{4} \sigma^2\vec{k}^2 - \mathrm{i}\vec{k}\cdot\vec{q}_1\bigg\} .	\nonumber
\end{align}
The remaining integral is actually divergent since the integrand is not decaying fast enough for $|q|\rightarrow\infty$, which is due to the infinite range of the Newtonian gravitational potential. To perform this Fourier transform properly, we thus introduce an infrared cutoff to regularize the integral, which finally leads us to
\begin{align}
	v_\mathrm{g}(\vec{k}) &=  GM A^2B \bigg(2\pi^2\sigma^4 \bigg)^{3/2} \int \mathrm{d}^3 q_1\  \frac{1}{|\vec{q}_1|}  \exp\bigg\{-\epsilon |\vec{q}_1|-\frac{3\sigma^2}{4}\vec{k}^2 - \mathrm{i}\vec{k}\cdot\vec{q}_1\bigg\} \label{MPH: v_g in static ST}\\
	&= - \frac{C_\mathrm{g}}{\vec{k}^2+\epsilon^2}\exp\bigg\{-\frac{3}{4}\sigma^2\vec{k}^2\bigg\}. 	\nonumber
\end{align}

%%%%%%%%%%%%%%%%%%%%%%%%%%%%%%%%%%%%%%%%
\section{Explicit Expressions for Free 2-Point Cumulants}
\label{Free Cumulants Appendix}
The calculation of the free cumulants is described in detail in \cite{fabis_kinetic_2018}. In lowest order in the initial power spectrum \cor{$P_\delta^{(\mathrm{i})}$, using growing mode initial conditions $\delta^{(\mathrm{i})} = - \vec{\nabla} \cdot \vec{p}^{\;(\mathrm{i})}$,} and ignoring shot-noise contributions, the free 2-point cumulants for our cosmological framework are given by
\begin{align}
	G_{BB}^{(0)}(1,2) &= 0, \label{MPH: G_0,BB} \\
	\cor{G_{nB}^{(0)}(1,2) = G_{Bn}^{(0)}(2,1)} &\cor{= -i \, (2\pi)^3 \delta_\mathrm{D}(\vec{k}_1+\vec{k}_2) \bar{n} \, \vec{k}_1^2 \, g_{12} ,} \label{MPH: G_0,nB} \\
	G_{nn}^{(0)}(1,2) &= (2\pi)^3 \delta_\mathrm{D}(\vec{k}_1+\vec{k}_2)  \bar{n}^2 P_\delta^{(\mathrm{i})}(\vec{k}_1)(1+g_1)(1+g_2), \label{MPH: G_0,nn}
\end{align}
where we have used the abbreviations 
\begin{align}
	g_{jk} &\coloneqq g_{qp}(\eta_j,\eta_k)\theta(\eta_j-\eta_k) ,   \\
	g_j &\coloneqq g_{qp}(\eta_j,0) ,
\end{align}
\cor{with $g_{qp}$ given in \eqref{qp-component of retarded Green's function},} and $P_\delta^{(\mathrm{i})}(k)$ denotes the power spectrum at initial time. Note that the contributions of higher orders in the initial power spectrum can be safely neglected in the qualitative analysis presented in this paper, as they only \cor{affect scales below the particles' mean free-streaming length} \cite{lilow_resummed_2019}.

\bibliography{bibliography}

\end{document}